\documentclass[pre,twocolumn,superscriptaddress,showpacs,preprintnumbers]{revtex4}

\newcommand{\bq}{\begin{equation}}
\newcommand{\ba}{\begin{eqnarray}}
\newcommand{\eq}{\end{equation}}
\newcommand{\ea}{\end{eqnarray}}

\newcommand{\calA}{\mathcal{A}}

\newcommand{\calD}{\mathcal{D}}

\newcommand{\calN}{\mathcal{N}}

\newcommand{\Det}[1]{\det [ \, #1 \, ]}
\newcommand{\Exp}[1]{\exp \{ \, #1 \, \} }
\newcommand{\bnabla}{\boldsymbol{\nabla}}

\newcommand{\Qquad}[1]{\qquad\text{#1}\qquad}      

\newcommand{\bv}{\mathbf{v}}
\newcommand{\bx}{\mathbf{x}}
\newcommand{\bk}{\mathbf{k}}

\newcommand{\LnB}[1]{\ln \Bigl \lbrack \, #1 \, \Bigr \rbrack }

\newcommand{\rd}{\mathrm{d}}
\newcommand{\Tr}[1]{\mathrm{Tr} [ \, #1 \, ]}
\newcommand{\Ln}[1]{\ln [ \, #1 \, ]}

\usepackage{amsfonts}
\usepackage{amssymb}
\usepackage{amsmath}
\usepackage{feynmp}
\usepackage{color}
\usepackage{graphicx}
\begin{document}
\title{Auxiliary Field Loop Expansion for the Effective Action for Stochastic Partial Differential Equations I}
\author{Fred Cooper} \email{cooper@santafe.edu}
\affiliation{Department of Earth and Planetary Science, Harvard University,Cambridge, MA 02138}
\affiliation{The Santa Fe Institute, 1399 Hyde Park Road, Santa Fe, NM 87501, USA}
%

\date{\today}

\begin{abstract}
Using a path integral formulation for correlation functions of  stochastic partial differential equations based on the Onsager-Machlup approach, we show how, by introducing a  composite auxiliary field one can generate an auxiliary field loop expansion for the correlation functions which is similar to the one used in the  $1/N$ expansion for an  $O(N)$  scalar quantum field theory. 
We apply this formalism to the Kardar Parisi Zhang (KPZ)  equation, and introduce the composite field
 $\sigma  =  \frac{\lambda}{2}   \nabla \phi \cdot  \nabla  \phi$ by inserting a representation of the unit operator into the path integral which enforces this constraint. In leading order we obtain a self-consistent mean field approximation for the effective action  similar to that used for the Bardeen-Cooper-Schrieffer (BCS) and Bose-Einstein Condensate (BEC) theories of dilute Fermi and Bose gases. This approximation, though related to a  self-consistent Gaussian approximation, preserves all symmetries and broken symmetries.  We derive the leading order in the auxiliary field (LOAF) effective potential and compare our results to the  one loop in the fluctuation strength $\calA$  approximation. We find, contrary to what is found in  the one loop and  self-consistent Gaussian approximation schemes that in the LOAF approximation there is no fluctuation induced symmetry breaking  as a function of the coupling constant in any dimension $d$.  \end{abstract}
\pacs{ 02.50.Ey , 05.10.Gg, 05.45-a}
\maketitle

\section{Introduction}

The effective action $\Gamma[\phi]$  for stochastic partial differential equations extends the role played by the  action for field theories of  classical dynamical systems.  
The effective action accounts for the physics of the entire system, composed of the various degrees of freedom (represented by the fields $\phi$) as well as the effect on them by stochastic agents in the form of noise.The first derivative of $\Gamma[\phi]$  with respect to the field gives the equation for the evolution of the field averaged over different realizations of the noise chosen from a given probability distribution. Higher derivatives of the action determine the one particle-irreducible vertices, from which all the correlation functions of the field can be reconstructed. The one-particle irreducible vertices play a crucial role in identifying the eventual renormalization of the parameters found in the theory without noise (or in quantum field theory without quantum fluctuations).  Most text books on quantum field theory (see for example \cite{Itzykson})   discuss the effective action and usually describe the semi-classical approximation which is an approximation which  keeps only the Gaussian fluctuations  around the classical motion.  In the recent literature most discussions of the effective action for stochastic partial differential equations  are based on a loop expansion in terms of the strength of the noise correlation function which is the same as the semi-classical approximation. For the KPZ equation, a recent discusson is found in \cite{Hochberg}.  In population biology, a recent discussion is found in \cite{Dodd} and in pair annihilation and for  Gribov processes a recent discussion is found in  \cite{Zorzano}. In the semi-classical approximation it is tacitly assumed that the noise is a small perturbation on the classical dynamics and that therefore perturbation theory is valid.  This separation is often not present in many dynamical systems acted upon by noise (whether internal or external). This shortcoming of the semi-classical approximation was seen even at small couplings in the theory of dilute Bose gases where the fluctuations are thermal in nature. In that situation the semiclassical approximation did not give  a true picture of the phase structure  \cite{Andersen}.   Because of this, in this and a subsequent  paper  \cite{us} we will develop a different approach \cite{BCG}  that is non-perturbative in both coupling constant strength and the strength of the noise correlation functions. This approach is based on introducing auxiliary fields which render the theory quadratic (Gaussian) in the original fields, without sacrificing the possibility of determining in a systematic way the corrections to the lowest order approximation which is a "self consistent" Gaussian approximation (but {\it not} a Gaussian truncation).  
This method  has been successfully used to determine approximately, in leading order,  the phase structure of dilute Bose \cite{BEC1} \cite{BEC2}  and Fermi gases \cite{BCS}.  
In general, the value of the effective action for constant fields , (divided by the space time volume) called the effective potential,  plays a role similar to the usual potential of Classical Mechanics as it maps the energy landscape of a system.

   The use of auxiliary fields in Many-Body theory and quantum field theory  has a long history starting with Hubbard and Stratonovich \cite{Hubbard}.  The auxiliary fields render the Classical action quadratic in the original fields in the path integral formulation of the theory. Because of this one can reproduce many self consistent Gaussian fluctuation approximations by  evaluating the (path)  integral over the auxiliary fields by steepest descent. In quantum field theory and many-body field theory applications, the leading approximation (LOAF)  determines the  large-N behavior of 
the O(N) model  \cite{CJP} , BCS theory of superconductivity  \cite{BCS}  and the  LOAF  theory for Bose-Einstein condensation \cite{BEC2}. The path integral approach allows   a completes reorganization of the Feynman graphs of the theory in terms of the self consistently obtained propagators for the original fields, and the leading order propagator for the auxiliary field \cite{BCG}. 
When the original theory starts from a quartic self interaction described by a coupling $\lambda$, the auxiliary fields that are utilized are quadratic in the original fields and the leading order point  self interaction gets replaced by a trilinear interaction involving two of the original fields and the auxiliary field. The scattering in lowest order  in the auxiliary field loop expansion proceeds by the intermediary of the composite field propagator.  The second derivative of the effective potential with respect to the auxiliary field gives the value of the inverse of the composite field propagator, and due to the constraint equation satisfied by the auxiliary field, one finds by studying the renormalized Green's functions that this is a renormalization invariant. This invariant in leading order is the inverse of the renormalized coupling constant \cite{SD1}.  Thus from this second derivative of the potential one can easily derive the renormalization group equation for the renormalized coupling constant. 

Another property of the auxiliary field loop expansion is that order by order it preserves exact symmetries and broken symmetries such as the Goldstone theorem \cite{Goldstone}.  Thus it is more trustworthy in determining broken symmetry phase transitions than other self-consistent approximations such as the Hartree-Fock-Bogoliuibov (HFB) approximations.
Self consistent  methods such as HFB often predict phase transitions when they are absent, or wrong order phase transitions \cite{Andersen}.  In our study of the KPZ equation we will find that our result for the phase structure  in leading order surprisingly gives different results from the usually used loop expansion (in the strength of fluctuations)  even at small values of the fluctuation correlation strength. In distinction to this semi-classical approximation and the self consistent Hartree approximation, we find in leading order (LOAF) that there is no phase transition in any dimension as a function of coupling constant. 

 In this first of two papers we will give the formulation for obtaining the effective action  based on the ideas of Onsager and Machlup \cite{Onsager} \cite{Graham} \cite{Zinn-Justin}.  In a follow-up paper  \cite{us} we will give a second formulation which is based on the response-theory formalism of Martin-Siggia-Rose  \cite{msr} \cite{Peliti} \cite{janssen} \cite{Jouvet}.
Although we concentrate on the static properties of the fields in these two papers, the formalism presented here  is perfectly well adapted to doing simulations for the time evolution of the field (averaged over noise configurations) as well as the noise induced  correlation functions. This approach has been used in the past to study the dynamics of phase transitions when there is chiral symmetry breaking \cite{chiral} , or phase separation in Bose Einstein condensates \cite{Chien}.

The paper is organized as follows.  In section II  we review  the path integral approach to stochastic partial differential equations of the reaction diffusion  type.
 In section III we review the auxiliary field loop expansion method which has been used successfully in understanding BCS theory \cite{BCS}  and BEC  \cite{BEC2}  theory as well as relativistic quantum field theories such as the scalar O(N) model  \cite{CJP} \cite{BCG} \cite{Nreview}. In section IV we review the KPZ equation and rewrite the action in terms of an auxiliary field 
 $\sigma$ which renders the theory quadratic in the initial fields $\phi$. We then we apply the auxiliary field loop expansion method to this problem and derive the first two terms  in the loop expansion for the effective action.  In section V we derive the effective potential  and discuss the LOAF result for renormalized effective potential for the critical dimension $d=2$.  In section VI we derive the renormalized  effective potential for $d=1$ and $d=3$. In section VII we discuss the renormalization group flows for the coupling constant for arbitrary $d$  found in this approximation.  In section VIII we summarize our results.   We leave to the appendix the details of how to perform the renormalization program using explicit cutoffs. %
\section{ Path integral formulation for Reaction diffusion equations with noise}
In this section we briefly review the path integral formulation for the field  correlation functions induced by external noise that has Gaussian correlations.  This formulation was originally developed by Onsager and Machlup \cite{Onsager} and later by Graham   \cite{Graham} and  Zinn-Justin \cite{Zinn-Justin}.  More recently it has been reviewed and elaborated on by Hochberg, Molina-Paris, Perez-Mercader and Visser \cite{Hochberg}.  We will follow the  approach of \cite{Hochberg}. 

A generic system of coupled reaction diffusion equations with external noise can be  written in the schematic   form:
\bq
\frac{\partial {\Phi_i}}{\partial t} - \nu_i \nabla^2  \Phi_i - F_i[\Phi] - \eta_i \equiv D_i \Phi_i - F_i[\Phi] - \eta_i = 0.  \label{phii}
\eq
We assume the noise is Gaussian in that the probability distribution function for noise can be described by 
\bq
P[\eta[x,t]] = N \exp \left[ - \frac{1}{2} \int dx dy  \eta^i(x)  {G_\eta} ^{-1}_{ij} (x,y)   \eta^j(y) \right] ,
\eq
where $N$ is determined from
\bq
\int \prod_i {\cal D} \eta_i P[\eta]  = 1.
\eq
For this distribution
\bq
\langle \eta^i (x,t) \rangle  \equiv \int \calD \eta ~ P[\eta[x,t]] ~  \eta^i (x,t) =  0 ,
\eq
and the two point noise   correlation function (connected Green's function) is given by 
 \bq
\langle \eta ^i (x,t)   \eta ^j (y,t') \rangle_c   =   G_\eta  ^{ij}(xt; y t').
\eq
We will assume that the strength of the noise correlation function is proportional to $\cal A$,  which is a parameter often used to control various approximation schemes.
In particular in our KPZ  example we will choose 
\begin{equation}
G_\eta  ^{ij}(xt; y t') = {\cal A } \delta^d (x-x') \delta(t-t').  
\end{equation}
  One assumes that for a particular configuration of the noise $\eta$, one can solve Eqs \ref{phii}
 for  $\Phi_i(x | \eta)$ and that there is a unique solution.  A strategy for doing this in the strong coupling domain is discussed in 
 \cite{bcf}.

The expectation values of the concentrations $ \Phi_i $ are obtained by performing the stochastic average over the noise

\bq \label{corr2}
\langle \Phi_i(x,t) \Phi_j(y,t') \rangle_{\eta}  = \int \calD  \eta P[\eta] \Phi_i(x,t | \eta) \Phi_j(y,t' | \eta ).
\eq

We are interested in getting a path integral representation for the correlation functions of Eq. (\ref{corr2}). 
Using the Fadeev-Popov trick we can, instead of explicitly solving for the $\Phi_i$ in terms of the noise $\eta$, enforce the fact that we are doing the integral over fields that obey the noisy reaction diffusion equation.  That is we insert the functional delta function into the path integral over $\eta$ using the identity:
\ba \label{delta}
&&1= \int \calD \Phi_i  \delta [\Phi_i - \Phi_i (x| \eta)] \nonumber \\
&&=\int \calD \Phi_i  \delta [\partial {\Phi_i}{\partial t} - \nu_i \nabla^2  \Phi_i - F_i[\Phi] - \eta_i]  | {\rm det}  S^{-1}[\Phi] |. \nonumber \\
\ea
Here 
\bq 
S^{-1}_{ij}  =  \left( D_i \delta_{ij}  -  \frac {\partial F_i[\Phi] }{\partial \phi_j } \right) \delta(x-y) \delta(t-t').
\eq
The determinant can be replaced by a path integral over fermionic fields or ignored if we use an appropriate choice of the
lattice version of the time derivative (forward derivative) which is the Ito regularization  \cite{ref:GN} \cite{Kamenev}.  We will assume the Ito regularization in what follows. 
At this point there are two approaches to obtaining a field theory description.  One can directly integrate over the noise to obtain
the approach of Onsager and Machlup \cite{Onsager}  \cite{Graham} \cite{Zinn-Justin}.   This is the approach we will follow here for the  Kardar-Parisi-Zhang equation \cite{kpz}. This will enable us to compare our findings with previous calculations of the effective potential by Hochberg et. al.  \cite{physica} and Amaral and Roditi \cite{Gaussian}. 
An alternative approach that also gives a path integral representation for the correlation functions would be to  introduce a functional representation for the delta function to obtain a Martin-Siggia-Rose  action \cite{msr}  \cite{janssen} \cite{Jouvet} \cite{Kamenev} which is closely related to the Schwinger-Keldysh formalism \cite{SK}.  We will dicuss that strategy in a subsequent paper \cite{us}.

The generating functional  $Z[j]$ for the correlation functions is given by
\bq
Z[j] =\int \calD \eta  P[\eta] \exp[\int  dx J_i(x) \Phi_i (x | \eta) ].
\eq
If we introduce the  delta function identity  Eq. \ref{delta} and performing the $\eta$ integrals we obtain the Onsager-Machlup formulation
for the generating functional:
\ba
Z[j] &&= \int \calD \Phi_i  P[ D \Phi_i - F_i[\Phi] ]   \exp[\int  dx J_i(x) \Phi_i (x | \eta) ]  \times \nonumber \\
&&   |det S^{-1}| ,
\ea 
where 
\bq
S^{-1}_{ij}(x,y)   = (\partial_t - \nu_i  \nabla^2) \delta (x-y) \delta_{ij} - \frac{ \delta  F_i[ \Phi(x])} {\delta{\Phi_j (y)}}.
\eq
This has been discussed in detail in \cite{ref:Hochberg1} and \cite{Gaussian}. The generating functional for the correlation function can be written as \cite{Hochberg}
\ba
&&Z[J] = \frac{1}{\sqrt{det (2 \pi S_\eta)}} \int \calD \phi \sqrt{K K^\dag} \exp \left( J \phi \right) \nonumber \\
&& \times \exp \left( - \frac{1}{2} \int (D \phi - F[\phi]) G_\eta ^{-1} (D \phi - F[\phi]) \right) ,\label{gen1} \nonumber \\
\ea
where 
\bq
K= det( D- \frac{\partial F} {\partial \phi}) ,
\eq
and we have introduced the notation
\begin{equation}\label{Ddagdefs}
   D
   =
   \frac{\partial}{\partial t}
   - 
   \nu \, \nabla^2 \>,
   \qquad
   D^{\dag}
   =
   -
   \frac{\partial}{\partial t}
   - 
   \nu \, \nabla^2 \>.
\end{equation}

\section{Auxiliary field loop expansion}
Now let us imagine that by means of introducing auxiliary fields $\sigma_k$   the action is now rendered quadratic in the original fields $\phi_i$.    We will make this specific for the KPZ equation below, but the reader is referred to the many examples of using this approach using  a Hubbard-Stratonovich transformation
\cite{Hubbard,BCG,CJP},  or in the large-N literature, a functional delta function defining the auxiliary field \cite{Nreview}.   After doing this the ``bare"  inverse propagator for the $\phi$ fields can be symbolically written as 
$G_\phi  ^{-1}[\sigma] $ 
and the full action can be written as 
\ba
S [\phi, \sigma] &&= \int  dx dy  dt dt' \left[  \phi_i[x,t, ]{ G^{-1}_\phi}  ^{ij}  (x,y, t, t'; \sigma)  \phi_j [y,t']  \right. \nonumber \\
&& \left. + S_2 \left[\sigma_i \right] \right] -  \int dx dt [j_i \phi_i+ J_k \sigma_k ] ,
\ea
where now $S_2[\sigma]$ is only a function of the auxiliary  fields $\sigma_i$.

After introducing the auxiliary fields, the generating function for the noise induced interactions (see Eq. \eqref{gen1})  is  now of the form:
\bq
Z[{J,j} ] = \int \calD \sigma  \calD \phi  \exp \left[ - { S}[\phi, \sigma  ] \right].  \label{gen2}
\eq

One can now perform the Gaussian Path  integral over $\phi$ to obtain an equivalent  action ${\hat S}$  which just depends on the field $\sigma$ and the external 
sources $J, j$. 
Performing the Gaussian integral, we are left with the  expression for the generating functional:
\ba
Z[{J,j}]&& = \int \calD \sigma  \exp [ - {\hat S}[\sigma,j, J]  ] ,  \label{seff1}
\ea
where
\ba
{\hat S} && =  \int dx  S_2[\sigma] -  \int dx dy \frac{1}{2}  j(x)   G(x,y; \sigma) j(y)  \nonumber \\
&&- \int dx J(x) \sigma(x)  +  \frac{1}{2} ~ {\rm Tr} ~\ln G^{-1} (\sigma)    \nonumber \\  \label{seff2a}
\ea
and we have added a source term $J$  for  the auxiliary field $ \sigma$.  $W[J, j] = \ln Z[j, J] ] $ is the generator of the connected correlation functions.

We next introduce a small parameter $\epsilon$  into the theory  via the substitution
${\hat S} \rightarrow {\hat S}/\epsilon$.  For small $\epsilon$ evaluation of the integral by steepest descent (or Laplace's method)  is justified. $\epsilon$ is similar to $\hbar$ in that it counts loops but now the loops are in the propagators for the auxiliary fields $\sigma$. 
The auxiliary field loop expansion is obtained by first  expanding  around the stationary phase point, and using the Gaussian term for the
measure of the remaining integrals expanded as a power series in $\epsilon$ \cite{BCG}. That is 
\ba  \label{ea.e:Sexpand} 
   &&{\hat S}[\, \sigma,J \,]
   =
   {\hat S}[\, \sigma_0,J \,] \nonumber \\
    &&  +
   \int [\rd x] \,
   \frac{\delta {\hat S}[\, \sigma,J \,]}
        {\delta \sigma^i(x)} \Big |_{\sigma_0}     ( \, \sigma^i(x) - \sigma_0^{i}(x) \, )
  \nonumber \\
  && 
   +
   \frac{1}{2} 
   \iint [\rd x] \, [\rd x'] \,
   \frac{\delta^2 {\hat S}[\, \sigma,J \,]}
        {\delta \sigma^i(x) \, \delta \sigma^j(x')} \Big |_{\sigma_0}    \times \nonumber \\
       && ( \, \sigma^i(x) - \sigma_{0}^{i}(x) \, ) \,
   ( \, \sigma^j(x') - \sigma_{0}^{j}(x') \, ) \nonumber \\
  && +
   \dotsb
   \notag
\ea
 The vanishing of the first derivatives define the saddle point $\sigma_0^i$, 
 \bq
\frac{\partial {\hat S}} {\partial \sigma_i} |_{\sigma=\sigma_0 ^i }   = 0,
\eq
leading to the ``gap" equations at the stationary phase point
\bq
\sigma_0^i = \sigma_0^i  [ \phi_c, \nabla \phi_c] >.
\eq 
Here
\bq
\phi_c (x) = < \phi(x) > = \frac {\delta \ln Z} {\delta j(x) } = \int dy G_\phi (x,y) j(y)
\eq
which corresponds to the equations of motion for the densities averaged over the noise in the presence of sources:
\bq
\int  dy G_\phi ^{-1}(x,y) \phi_c(y) = j (x).
\eq
Keeping terms up to quadratic in the expansion about the stationary phase point and performing the Gaussian integral in $\sigma$   we obtain for the $W[J,j] $

\bq
-W[J,j ] =  {\hat S}[\, \sigma_0,J \,] + \frac{\epsilon}{2} \Tr {\ln D_\sigma ^{-1} [\sigma_0]} ,
\eq
where 
\bq \label{Dinv1}
 D_{ij}^{-1} [\sigma_0](x,x') = \frac{\delta^2 {\hat S}[\, \sigma,J \,]}
        {\delta \sigma^i(x) \, \delta \sigma^j(x')} \Big |_{\sigma_0} 
\eq
is the inverse propagator for the $\sigma$ field evaluated at the stationary phase point. 
Higher terms in the loop expansion in the  $\sigma$ propagator are obtained by treating the higher terms in the derivative  expansion of $ {\hat S}[\sigma]$ perturbatively with respect to the Gaussian measure.  This is discussed in detail in \cite{BCG}. 
 $\Gamma [ \phi, \sigma ]$, which is the generator of the one-particle irreducible graphs,  is the Legendre Transform of  $Z[j, J]$, i.e.
\bq
\Gamma [\phi, \sigma] =  -\ln Z[J,j ] + \int  dx [  j  \phi+J \sigma] ,
\eq
where now
 \bq
\phi = < \phi>  =  \frac{ \delta  \ln Z[J] } {\delta j} ; ~~ \frac{ \delta \Gamma [ \phi, \sigma] }{\delta \phi} = j.
\eq
\bq
\sigma = < \sigma>  =  \frac{ \delta  \ln Z[J,j ] } {\delta J} ; ~~ \frac{ \delta \Gamma [ \phi,  \sigma] }{\delta \sigma} = J.
\eq

Again if we keep the stationary phase point plus Gaussian fluctuations we obtain schematically for $\Gamma$ up to order $\epsilon$
\ba  \label{effact1}
\Gamma [\phi, \sigma] && = \frac{1}{2}  \int  dx dy  dt dt' \left[  \phi[x,t, ] G_\phi^{-1}(x,y, t, t'; \sigma)  \phi[y,t']  \right. \nonumber \\
&& \left. + S_2 [\sigma]  \right] 
 + \frac{1}{2} \Tr {\ln G_\phi^{-1} [\sigma ]} \nonumber \\
&&  +  \frac{\epsilon}{2} \Tr {\ln D_\sigma^{-1} [\sigma, \phi]}.
\ea
The LOAF approximation consists of just keeping the stationary phase part of the effective action (i.e. $\epsilon \rightarrow  0$).  The effective potential $V_{eff}$
is the value of $\Gamma$ for constant fields divided by the Volume.  $V_{eff}$ is similar to the usual potential energy in Classical mechanics as it describes the energy landscape.

\section{KPZ equation}
At this point we need to be specific about the reaction-diffusion equation so that we can introduce auxiliary fields that are appropriate to the dynamics of the problem.  Since the KPZ equation has been considered before in discussions of the effective action, we will consider in this first paper, the LOAF approximation for the KPZ equation.
In a subsequent paper  \cite{us} we will study the LOAF approximation for both the annihilation process $A+A \rightarrow 0$ and the Ginzburg Landau model for spin relaxation in the related MSR formalism. 
The idea of using the  Effective Potential to discuss the phase structure of the massless KPZ was emphasized in the work of Hochberg et.al. 
\cite{physica},  who evaluated the effective potential in a loop expansion in the noise strength parameter. In that paper one found that there was spontaneous symmetry breakdown caused by the fluctuations in one and two dimensions.  The approach of \cite{physica}   was generalized to a self-consistent Gaussian approximation by Amaral and Roditi \cite{Gaussian}. In their study spontaneous symmetry breakdown was found not only in $d=1,2$  also in $d=3$.   In the theory of Bose Einstein condensates (BECs), the loop expansion is what is called the Bogoliubov approximation, and the self consistent Gaussian approximation is what is known as the Hartree-Fock-Bogoliubov approximation (see the review in \cite{Andersen}).    In the theory of Bose Einstein condensation, the presence of a broken symmetry which leads to the existence of a Goldstone boson is crucial to the understanding of the phase structure.  In that situation the HFB approximation gave misleading answers as a result of violating the Goldstone theorem.  On the other hand, an auxiliary field loop expansion which in leading order (LOAF) was similar to the HFB approximation, but  preserved the Goldstone theorem, predicted the correct order of phase transition to the BEC phase \cite{BEC1}.  Thus the results of our LOAF prediction for the KPZ effective potential, presented below, which shows that there is no fluctuation induced symmetry breakdown casts doubts on some of the conclusions of the previous calculations.

 For the KPZ equation, the physical ``gauge invariant" degree of freedom is $\nabla \phi$  which plays the role of 
an electric field.  For the KPZ equation
\bq
F[\phi] = f_0 +  \frac {\lambda^2 }{2}  (\nabla \phi)^2.
\eq
Here $f_0$ is a ``tadpole" source that is needed for the normalization process as will be seen.  Its renormalized version will be set equal to zero so that the infrared properties of the theory will be that of the massless KPZ equation. 
For noise sources whose correlation functions  are translation invariant and local in time the KPZ equation has the Galilean symmetry:
\ba
&& {\vec x'} ={\vec x} +  \lambda  {\vec v} t, \nonumber \\
&&t = t', \nonumber \\
&&  \phi'(x,t)= \phi(x,t) - {\vec v} \cdot {\vec x} + \frac{\lambda}{2} v ^2 t.
\ea

We want to   introduce a new composite auxiliary field $\sigma / \lambda= f_0+  \frac {\lambda }{2}  (\nabla \phi)^2 $, to mimic large-$N$  expansion methods used in Bosonic field   theories with quartic self-interactions \cite{CJP} \cite{Nreview} \cite{BEC1}.  This can be done by introducing another functional delta function into the path integral
\ba
&&1 = \int d \sigma  \delta( \frac{\sigma }{\lambda}-   \frac {\lambda }{2}  (\nabla \phi)^2 - f_0) \nonumber \\
&& = \int ~ d  \sigma ~  d \chi \exp[\int dx \frac{\chi}{\calA \lambda } ( \frac{\sigma }{\lambda}- f_0-  \frac {\lambda }{2}  (\nabla \phi)^2) ]. \label{fdelta}
\ea
 In \eqref{fdelta} the integration is along the imaginary axis as discussed in \cite{Nreview}. Here $\calA$ is the strength of the noise term, introduced so that the entire action
 is proportional to $\calA^{-1}$. 
 
 An alternative method for introducing the auxiliary field $\sigma$ is the  Hubbard-Stratonovich transformation \cite{Hubbard}.  In that approach (which is equivalent to what we do here)  one would add to the original KPZ action a term of the form 
\ba
&&S_{HS} = \nonumber \\
&& \frac{1}{2 \calA}  \int dx dt [ ( \frac{\sigma }{\lambda}- f_0-  \frac {\lambda }{2}  (\nabla \phi)^2) G_\eta^{-1}  ( \frac{\sigma }{\lambda} - f_0-  \frac {\lambda }{2}  (\nabla \phi)^2). \nonumber \\
\ea

Once we insert the identity Eq. \eqref{fdelta}  
 into the path integral for the generating functional Eq. \eqref{gen1}   and introducing additional currents $s(x)$ and $r(x)$ for the auxiliary fields we then obtain
\ba \label{KPZ.e:Z-III}
  && Z[ \, j,s,r \, ]
   = \nonumber \\
   &&\frac{ 1 }{ \sqrt{ \Det{ 2\pi G_{\eta} } } }
   \iiint \calD \phi  \calD \sigma  \calD \chi 
   \Exp{ - S'[  \phi,\sigma,\chi;j,s,r  ] } \>. \nonumber \\
\ea
After integrating by parts and  discarding surface terms, the action $S'$ can be written as
\ba
   &&S'[  \phi,\sigma,\chi;j,s,r  ]
   = \nonumber \\
&&   \frac{1}{2} 
   \iint \rd x  \rd x' 
 \left[
      \phi(x) G_{\phi}^{-1}[\chi](x,x') \phi(x') \right. \nonumber \\
    && \left.   +
      \sigma(x) 
      G_{\eta}^{-1}(x,x') 
      \sigma(x') 
\right]  \nonumber  \\
   & &
   +
   \int \! \rd x \, 
 \left[ 
      \frac{\chi(x)} {\calA  \lambda}  \, ( \frac{\sigma(x)}{\lambda} - f_0 ) \right.  \nonumber \\
  && \left.    -
      J[ \, j,\sigma \, ](x) \, \phi(x) 
      -
      s(x) \, \sigma(x)
      -
      r(x) \, \chi(x) \,
 \right]  \nonumber \\
   \label{KPZ.e:Sp-II}
\ea
where
\begin{equation}\label{KPZ.e:Gphiinvedef}
   G_{\phi}^{-1}[\, \chi \,](x,x')
   =
   D^{\dag} \, D^{\prime\dag} \, G_{\eta}^{-1}(x,x')
   +
   \bnabla \cdot \bnabla' \, G^{-1}_{\chi}[\, \chi \,](x,x') \>
\end{equation}
 with $J[\, j,\sigma \, ](x)$ given by 
\begin{align}
   J[ \, j,\sigma \, ](x)  \label{KPZ.e:Jdef} 
   &=
    j(x)
   +
   D^{\dag} \int \rd x' \, 
   G_{\eta}^{-1}(x,x') \, \sigma(x') \>.
   \notag   
\end{align}
and 
\begin{equation}\label{KPZ.e:Gchiinvdef}
   G^{-1}_{\chi}[\chi](x',x)
   =
   \frac{\chi(x)} {\calA}  \, \delta(x-x') \>.  
\end{equation}

The action in \eqref{KPZ.e:Sp-II} is now quadratic in the field $\phi$  so that one can perform the path integration in the field  $\phi$.  
%
%
\section{\label{s:EffAction}Effective action}

A systematic expansion in auxiliary fields has been explained in the context of quantum field theory in previous work  \cite{BCG,BEC2}. .
Inserting the action $S'[ \, \phi,\sigma,\chi;j,s,r \, ]$ from \eqref{KPZ.e:Sp-II} into the path integral for the generating functional \eqref{KPZ.e:Z-III} and integrating over $\phi$ gives:
\ba \label{KPZ.e:Z-IV}
   Z[ j,s,r ]
  && =
   e^{ W[ \, j,s,r \, ] } \nonumber \\
  && =
   \calN
   \iint \calD \sigma \, \calD \chi \,
   \Exp{ - S_{\text{eff}}[ \, \sigma,\chi;j,s,r \, ] } \>, \nonumber \\
\ea
where the effective action is 
\ba 
  && S_{\text{eff}}[ \, \sigma,\chi;j,s,r \, ] = 
   \label{KPZ.e:Seff-I}   \nonumber \\
   && \frac{1}{2} 
   \iint \rd x \rd x' 
\left[ 
      J[\, j,\sigma \,](x) \, G_{\phi}[\, \chi \,](x,x') \, J[\, j,\sigma \,](x') \right.  \nonumber \\
  &&\left.     +
      \sigma(x) \,
      G_{\eta}^{-1}(x,x') \,
      \sigma(x') \,
\right] \nonumber    \\
  & & 
   +
   \int \! \rd x \, 
    \left[ 
      \frac{\chi(x)} {\calA  \lambda}  \, ( \frac{\sigma(x)}{\lambda} - f_0 )  \right. \nonumber \\
    &&\left.   -
      s(x) \, \sigma(x)
      -
      r(x) \, \chi(x)
      +
      \frac{1}{2} \, \Ln{ G_{\phi}^{-1}[\, \chi \,](x,x) } \,
\right]  \>. \nonumber \\
\ea
   Here we have ignored the Jacobian factor in the path integral.  The remaining path integral in \eqref{KPZ.e:Z-IV} is done by the method of steepest descent by inserting an $\epsilon^{-1}$ factor in front of the entire effective action. 
Note that the effective action contains terms of order $ \calA^{-1}$ , namely the ``classical piece" and the $\Tr { \ln  G_\phi^{-1}}  $ term which is of zeroth order.    The saddle point is at the solution of the equations,
\begin{subequations}\label{KZP.e:EAsaddlept}
\begin{align}
   \biggl [ \,
      \frac{ \delta S_{\text{eff}}[ \, \sigma,\chi;j,s,r \, ] }{\delta \chi(x)}
   \biggr ]_{\chi_0,\phi_0} \!\!
   &=
   0 \>,
   \label{KZP.e:EAsaddlept-a} \\
   \biggl [ \,
      \frac{ \delta S_{\text{eff}}[ \, \sigma,\chi;j,s,r \, ] }{\delta \sigma(x)}
   \biggr ]_{\chi_0,\phi_0} \!\!
   &=
   0 \>.
   \label{KZP.e:EAsaddlept-b}   
\end{align}
\end{subequations}
In terms of the ``classical''   field $\phi_0[J](x)$ defined  at the saddle point by
\begin{equation}\label{KPZ.e:phi0def}
   \phi_0[\, j,\chi,\sigma \,](x)
   =
   \int \rd x' \, G_{\phi}[\, \chi \,](x,x') \, J[\, j,\sigma \,](x') \>, 
\end{equation}
we obtain  for the  first saddle point equation Eq.~\eqref{KZP.e:EAsaddlept-a} \
\ba
 &&  \frac{ \sigma_0(x) }{ \lambda }
   = f_0 + 
   \frac{\lambda}{2} \, 
   \big | \,  \bnabla \, \phi_0[\, j,\chi_0,\sigma_0 \,](x) \, \big |^2
   +
   \lambda \calA  r(x)
   \label{KPZ.e:SPeq-1} \nonumber  \\
  &&
   - \calA
   \frac{\lambda}{2} 
  \bigl [ \,
      \bnabla_1 \cdot \bnabla_2 \,
      G_{\phi}[\, \chi_0 \,](x_1,x_2) \,
   \bigr ]  |_{x_1=x_2=x}\>.
   \nonumber \\
   \ea
Eq.~\eqref{KZP.e:EAsaddlept-b} leads to 
\ba \label{KPZ.e:EAsaddlept-b-II}
  && \frac{\chi_0(x)}{ \calA \lambda^2}
   = \nonumber \\
  && -
   \int \rd x' 
   G_{\eta}^{-1}(x,x') 
   \bigl [  
      \sigma_0(x')
      +
      D' \, \phi_0[\, j,\chi_0,\sigma_0 \,](x') \,
   \bigr ]
   +
   s(x) \>. \nonumber \\
\ea
Expanding $S_{\text{eff}}[ \, \sigma,\chi;j,s,r \, ]$ to second order in the auxiliary fields about the saddle point and performing the remaining path integral yields
\bq
- \ln Z[j, s,r]  = S_{eff} [ \chi_0, \phi_0 , \sigma_0 ]+  \epsilon \frac{1}{2} \ln [D^{-1} (\phi_0,\chi_0, \sigma_0) ] ,
\eq
where $D^{-1} (\phi_0,\chi_0, \sigma_0)$ is the matrix inverse propagator of the composite fields defined by the second derivatives of the action with respite $\chi,\sigma$ at the stationary phase point (see Eq. \eqref{Dinv1}.   

To the same order in $\epsilon$  the generator of the one particle irreducible graphs $\Gamma[\phi, \sigma, \chi] $ is given by 
\ba
\Gamma[ \, \phi,\sigma,\chi \, ] &&= S_{cl}[ \phi, \sigma, \chi] + \frac{1}{2} \Tr {\ln G^{-1} [\chi ]}   \nonumber \\
&&+ \epsilon  \frac{1}{2} \Tr {\ln D^{-1} [\sigma, \chi  \phi]}.
\ea
where 
\ba
 &&S_{cl} [ \, \phi,\sigma,\chi\, ]
   = \frac{1}{2} 
   \iint \rd x \, \rd x' \,
   \left[  
      \phi(x) \, G_{\phi}^{-1}[\, \chi \,](x,x') \, \phi(x')  \right. \nonumber  \\
    && \left.   +
      \sigma(x) 
      G_{\eta}^{-1}(x,x') \,
      \sigma(x')        -
      \left.  \bigl [ \, D \, \phi(x) \, \right] 
      G_{\eta}^{-1}(x,x') \,
      \sigma(x') \,
\right] \nonumber \\
   &&   +
   \int \! \rd x \, 
   \Bigl \{ \,
      \frac{\chi(x) } {\calA \lambda} \, (\frac {\sigma(x)}{\lambda} - f_0 )   \Bigr \}  \nonumber \\
    \ea

To leading order, which we call the LOAF approximation, we have
\bq
\Gamma[ \, \phi,\sigma,\chi \, ] = S_{cl}[ \phi, \sigma, \chi] + \frac{1}{2} \Tr {\ln G^{-1} [\chi ]}. \label{gamloaf}
\eq

Since both $ G_{\phi}^{-1}(x,x') $ and $G_{\eta}^{-1}(x,x') $ are proportional to $\calA^{-1}$, the last term in Eq. \eqref{gamloaf}   is  one order higher in
 $\calA$ but treated at the same order in the loop expansion parameter  $\epsilon$.  
 
  Now that we have the effective action we can write down the equations for the time evolution of the correlation functions.  For example the equation for $\langle \phi (x,t) \rangle_\eta  $ is now just given by 
 \bq
\frac{ \delta \Gamma[ \, \phi,\sigma,\chi \, ] } {\delta \phi(x,t) } = 0.
\eq
To solve this one must first eliminate the constraint field $\chi$, and also solve for the auxiliary field $\sigma$ in terms of $\phi$ and its correlation function.  Examples for studying studying the LOAF time evolution equations for phase transitions  (which are similar to the time dependent Hartree-Fock equations)  are found in \cite{chiral} and \cite{Chien}. 
 
%
%
\subsection{\label{ss:EffPot}Effective potential}

To make contact with previous work on the KPZ effective potential, we now  restrict ourselves to  a white noise source with amplitude $\calA$, so that
\begin{equation}\label{KPZ.e:whitenoise}
   G_{\eta}(x,x')
   =
   \calA \, \delta(x - x') \>,
 G_{\eta}^{-1}(x,x')
   =
   \delta(x - x') / \calA \>.  
\end{equation}
The effective potential is obtained by choosing the  auxiliary fields $\sigma$ and $\chi$ to be  constant in space and time. When we turn off the sources we are interested in  physically relevant field configurations $\phi_0(x)$  which are of the form $\phi_0(x) = - \bv \cdot \bx$.
 in line with the Galilean symmetry. 
 The latter definition is consistent with $\bv = - \bnabla \, \phi_0(x)$ being a constant velocity field.  For this field configuration   $D \, \phi_0(x) = 0$.   
For static fields $ \chi,  v, \sigma $, we can Fourier transform $G_\phi ^{-1}$ and determine the effective potential, which is now defined as 
\bq
  V_{\text{eff}}[ \, v, \sigma, \chi \, ]
   =
   \frac{ \calA \, \Gamma[ \, - \bv \cdot \bx,\sigma,\chi \, ] }{ \Omega },   \label{veff}
 \eq 
where we have rescaled $V$ by $\calA$ so that the classical term is independent of $\calA$, and where $\Omega = \int \rd x$ is the space-time volume.

From \eqref{KPZ.e:Gphiinvedef}, and integrating by parts, we find that for $\phi(x) = \phi_0(x)$, 
\begin{align}
   & \frac{1}{2}
   \iint \rd x \, \rd x' \,
   \phi(x) \, G_{\phi}^{-1}(x,x') \, \phi(x')
   \label{KPZ.e:Vterm1} 
   =
   - \frac{ \chi \, v^2 }{ 2  \calA } \, \Omega \>.
   \notag
\end{align}

We also have
\begin{equation}\label{KPZ.e:Vterm2}
   \iint \rd x \, \rd x' \,
    \sigma(x) \, G_{\eta}^{-1}(x,x') \, \sigma(x')
    =
    \frac{ \sigma^2 }{ \calA } \, \Omega \>,
\end{equation}
and
\begin{equation}\label{KPZ.e:Vterm3}
   \int \rd x \,
   \frac{ \chi(x) }{\lambda} \, [ \, \frac{ \sigma(x)}{\lambda}  - f_0 \, ]
   =
     \frac{ \chi }{\lambda} \, [ \, \frac{ \sigma}{\lambda}  - f_0 \, ]\, \Omega \>.
\end{equation}
For the  $ {\rm Tr } \ln$  term, we start by noting that the $\delta$-function has the Fourier transform representation,
\begin{equation}\label{KPZ.e:deltaFT}
   \delta(x - x')
   =
   \int \rd k \> 
   e^{ i [ \, \bk \cdot (\bx - \bx' ) - \omega ( t - t' ) ] } \>, 
\end{equation}
where we have defined
\begin{equation}\label{KPZ.e:rdkdef-II}
   \int \rd k 
   \equiv
   \iint \frac{ \rd^d k \, \rd \omega }{ (2\pi)^d \, 2\pi } \>.
\end{equation}
Then $G_{\eta}^{-1}(x,x')$ and $G_{\chi}^{-1}(x,x')$ have the representations,
\begin{subequations}\label{KZP.e:GetaGchiExpand}
\begin{align}
   G_{\eta}^{-1}(x,x')
   &=
   \int \rd k \, \frac{ 1 }{ \calA } \,
   e^{ i [ \, \bk \cdot (\bx - \bx' ) - \omega ( t - t' ) ] } \>,
   \label{KZP.e:GetaGchiExpand-a} \\
   G_{\chi}^{-1}(x,x')
   &=
   \int \rd k \,  \frac{\chi }{\calA} \,
   e^{ i [ \, \bk \cdot (\bx - \bx' ) - \omega ( t - t' ) ] } \>.
   \label{KZP.e:GetaGchiExpand-b}
\end{align}
\end{subequations}
From \eqref{KPZ.e:Gphiinvedef}, $G_{\phi}^{-1}[\, \chi \,](x,x')$ has the representation,
\begin{align}
   G_{\phi}^{-1}[\, \chi \,](x,x')
     &=
   \int \rd k \>
   \tilde{G}_{\phi}^{-1}(k) \,
   e^{ i [ \, \bk \cdot (\bx - \bx' ) - \omega ( t - t' ) ] } \>,
   \notag   
\end{align}
where
\begin{equation}\label{KPZ.e:tGphiinv-I}
   \tilde{G}_{\phi}^{-1}(k)
   =
   \frac{1}{\calA} \,
  \left[ \, \omega^2 + \nu^2 \, k^4 \, 
   -
    \, \chi \, k^2 \right] \>.
\end{equation}
The  $ {\rm Tr} ~~ \Ln $  term in the effective potential is therefore
\begin{equation}\label{KPZ.e:TrLnresult}
   \frac{1}{2}   {\rm Tr}  \ln G_{\phi}^{-1}   =
   \frac{\Omega}{2} 
   \int \rd k \,
   \LnB{ 
       \,
      ( \, \omega^2 + \nu^2 \, k^4 \, 
      -
    \, \chi \, k^2 )\,
       } \>.
\end{equation}
The  effective potential  in the LOAF approximation is therefore
\ba
   V_{\text{eff}}[ \, v, \sigma, \chi \, ]
   &&=
   \frac{ \calA \, \Gamma[ \, - \bv \cdot \bx,\sigma,\chi \, ] }{ \Omega }
   \label{KPZ.e:Veff-I}  \nonumber \\
  & &=
   \frac{ \sigma^2 }{ 2 \lambda^2 }
   +
 \frac{ \chi }{\lambda} \,
   \Bigl [ \,
     \frac{ \sigma}{\lambda}  - f_0 - \frac{ \lambda \, v^2 }{ 2 } \, 
   \Bigr ]  \nonumber  \\
 &&  + 
   \frac{\calA}{2} 
   \int \rd k \,
   \LnB{  \,
      ( \, \omega^2 + \nu^2 \, k^4 \, 
      -
     \chi \, k^2) \,
       }  \>. \nonumber \\
\ea
Minimizing the effective potential with respect to $\sigma$ gives
\begin{equation}\label{KPZ.e:dVeffdsigma}
   \frac{ \partial V_{\text{eff}}[ \, v, \sigma, \chi \, ] }
        { \partial \sigma }
   =
   0 \>,
   \Qquad{$\Rightarrow$}
   \chi = - \sigma \>.
\end{equation}
Substituting this into \eqref{KPZ.e:Veff-I} to eliminate  the Lagrange multiplier field $\chi$ yields
\ba
&& V_{eff} [\sigma,v] =- \frac{\sigma^2}{2 \lambda^2}+ \frac{ \sigma}{\lambda} (\frac{\lambda v^2}{2} + f_0)  \nonumber \\
 && + \frac{\calA}{2} \int \frac {d^d k  d \omega} {(2 \pi)^{d+1}} \ln \left[ \omega^2+ \nu^2 k^4 + k^2 \sigma \right].  \label{veffloaf}
\ea
When $\calA =0$ and we replace $\sigma$ by its value at the minimum, we recover the ``classical'' potential for this problem.

We can next perform the integration over $\omega$ using the identity:
\bq
\int_{- \infty}^\infty ~d \omega \ln \left(\frac{\omega^2+X^2}{\omega^2+Y^2} \right) = 2 \pi (X-Y) ,
\eq
to obtain 
\bq
V_{eff} [\sigma, v] =  - \frac{\sigma^2}{2 \lambda^2} + \frac{\sigma}{\lambda} ( \frac{\lambda v^2}{2} + f_0) +  \frac{\calA \nu }{2} K[\frac{\sigma}{\nu^2}] ,
\eq
where
\bq
K[ m^2 ] =  \int  \frac{d^d k}{(2 \pi)^d} |k | \sqrt{k^2 +  m^2}.
\eq

The gap equation for sigma is then given by 
\bq
\frac {\partial V_{eff}}{\partial \sigma} =  \frac{-\sigma}{\lambda^2} + \frac{1}{\lambda} ( \frac{\lambda v^2}{2} + f_0) + 
\frac{\calA } {4 \nu} \int  |k|  \frac{d^d k}{(2 \pi)^d} \frac{1}{ \sqrt{ k^2 + \frac{\sigma}{\nu^2}}}=0.
\eq
One can solve for $\sigma$ as a function of $v$ using:
\bq
\sigma =  \lambda \left(  \frac{\lambda v^2}{2} + f_0 \right) + \frac{\calA  \lambda^2 } {4 \nu} \int  |k|  \frac{d^d k}{(2 \pi)^d} \frac{1}{ \sqrt{ k^2 + \frac{\sigma}{\nu^2}}}.
\eq
The second derivative of the effective potential is just the  negative of the  inverse propagator for the $\sigma$ field evaluated at zero momentum transfer. 
\bq
-D^{-1}_{\sigma \sigma}  =  \frac{ \partial ^2  V } {\partial \sigma ^2} =- \frac{1}{\lambda^2} -  \frac{\calA  } {8 \nu^3 } I_1[\sigma/\nu^2, d] 
\eq
where
\bq
I_1[m^2,d] =  \int  |k|  \frac{d^d k}{(2 \pi)^d} \frac{1}{ ( k^2 + m^2)^{3/2}}.
\eq
The integral has M.ultraviolet divergences for $d \ge 2$ and in less than two dimensions it has infrared divergences when $ \sigma \rightarrow 0$.  Therefore it is necessary to introduce a mass scale $\mu^2 \neq 0$  to define the coupling constant.  The running coupling constant in the theory defined in terms of the auxiliary field is just the value of the correlation function $D_{\sigma \sigma}$. 
\bq
\lambda_r^2[q, \omega, \mu^2] = D_{\sigma \sigma}  [q, \omega, \mu^2].
\eq
We can perform the integral in $d$ dimensions to obtain the dimensionally regulated result:
\bq 
 I_1 [m^2 ,d] = m^{d-2}   I[d],
 \eq 
 where
 \bq
 I[d] = \frac{2^{1-d} \pi ^{-\frac{d}{2}-\frac{1}{2}} \Gamma
   \left(1-\frac{d}{2}\right) \Gamma \left(\frac{d+1}{2}\right)}{\Gamma
   \left(\frac{d}{2}\right)}  \label{I1}.
  \eq
 Near the critical dimension $d=2$  it is useful to extract the pole and write
 \bq
 I[d] =  \frac{J[d]}{2-d}.
 \eq
When $d$ is near two, we have the expansion:
 \ba
&& m^{d-2} \Gamma[1-d/2] =  \nonumber \\
&& \frac{2}{2-d} \left[ 1-  \frac{(2-d)}{2} \left(\gamma_E+ \ln{\pi \mu^2}  \right) + O ((d-2)^2 ) \right]  \label{expand}. \nonumber \\
 \ea

 In one and three dimensions the dimensionally regulated answer for $I_1$ given by Eq. \eqref{I1} is finite. In two dimensions, rewriting the theory in terms of the renormalized coupling constant again renders the theory finite.  In the appendix we will address the renormalization process using a cutoff  for completeness.  When we introduce a momentum space cutoff $\Lambda$,  we will find that the effective potential has divergences as $\Lambda \rightarrow \infty$  related to the vacuum energy,  the  tadpole $f_0$ as well as  the coupling constant. 
  
We  introduce a  renormalized coupling constant and a reference mass $\mu^2$  by evaluating the second derivative of the potential at $\sigma/ \nu^2 = \mu^2$.
 We have
 \bq  \label{flowa}
-\frac{1}{\lambda_r^2 [\mu^2]} =  \frac{ \partial ^2  V } {\partial \sigma ^2}|_{\sigma=\mu^2 \nu^2}  =- \frac{1}{\lambda^2} -  \frac{\calA  } {8 \nu^3 } \mu^{d-2} I[d], 
\eq
or
\bq
\lambda_r^2[\mu^2] = \frac{\lambda^2}{ 1+  \frac{\calA \lambda^2  } {8 \nu^3 } \mu^{d-2} I[d]}.
\eq

Then we can rewrite the second derivative of the potential in the form
\bq \label{vpp}
-\frac{1}{\lambda_r^2 [\sigma/\nu^2]]} =  \frac{ \partial ^2  V } {\partial \sigma ^2}  =- \frac{1}{\lambda^2[\mu^2] } -  \frac{\calA  } {8 \nu^3 }I[d]  (m^{d-2} -  \mu^{d-2} ) ,
\eq
where here $m^2 =\sigma/\nu^2$. 
This expression is manifestly finite for $ d \neq 2$ and the subtraction renders it finite in $d=2$. If we now integrate this equation once and then twice, taking into account the form
of the classical action to determine the constants of integration, we will obtain the renormalized gap equation as well as the renormalized potential.

\subsection{couping constant  flows in the  LOAF approximation} 
From the effective potential we were able to determine  the auxiliary field  inverse propagator and therefore the dependence of $\lambda_r^2$ on the scale $\mu$ (see Eq. \eqref{flowa}).   By differentiating  $\lambda_r[\mu^2 ] $  with respect to $ \mu$ keeping the bare coupling constant fixed we obtain  $\beta_\lambda$.
\bq
\beta_\lambda = \mu \frac{ \partial \lambda_r^2[\mu^2]}{\partial \mu} =  \frac{\calA  } {8 \nu^3 } J[d] \lambda_r^4.  \label{betalam}
\eq
This is of course, equivalent to   Eq. \eqref{vpp} for the flow which can also 
be written in the form
\bq
\frac{1}{\lambda_r^2 [\mu^2]]} =  \frac{1}{\lambda^2[\mu_0^2] } +  \frac{\calA  } {8 \nu^3 }I[d]  (\mu^{d-2} -  \mu_0^{d-2} ) ,
\label{vppb}
\eq
which relates the value of the coupling at two different scales.  
The  coupling constant depends on $\calA$ and $\nu$ as well.  These also run with scale and in this approach we would have to calculate these renormalizations separately in perturbation theory \cite{kpz}  or by going to next order in the auxiliary field loop expansion.  Another approach to determine the running of $\nu$ and $\calA$ was taken by Zanella and Calzetta \cite{Zanella}, who apply the RG directly to the effective action calculated in perturbation theory to order $\lambda^2$.  The part of the RG equation that we have determined is $\beta_\lambda$.

A dimensionless coupling constant $g_0[\mu^2] $  can be  defined in terms of the bare coupling $\lambda^2$ by 
\bq
g_0 [\mu^2] =  \frac{\calA  \lambda^2  } {8 \nu^3 } \mu^{d-2} \equiv {\cal B} \lambda^2 \mu^{d-2}.
\eq
Similarly the dimensionless renormalized coupling  at $\mu^2$  is defined as 
\ba
g_r  [\mu^2] &&= {\cal B}  \lambda_r^2 [\mu^2]   \mu^{d-2} \equiv  g_0 [\mu^2] Z_g [\mu^2]  \nonumber \\
( Z_g )^{-1} &&= 1+ g_0  [\mu^2] I[d] = 1+ g_0  [\mu^2]\frac{ J[d] }{2-d}.
\ea
From this we can calculate the usual  $\beta$ function  (with the caveat that we have {\it not} included the running of $\nu$ and $\calA$), 
\bq
 \beta(g_r) = \mu \frac{ \partial g_r  [\mu^2] }{ \partial \mu} = (d-2) g_r  + J[d] g_r^2.
\eq

\begin{figure}
\begin{centering}
\includegraphics[width=0.8\columnwidth]{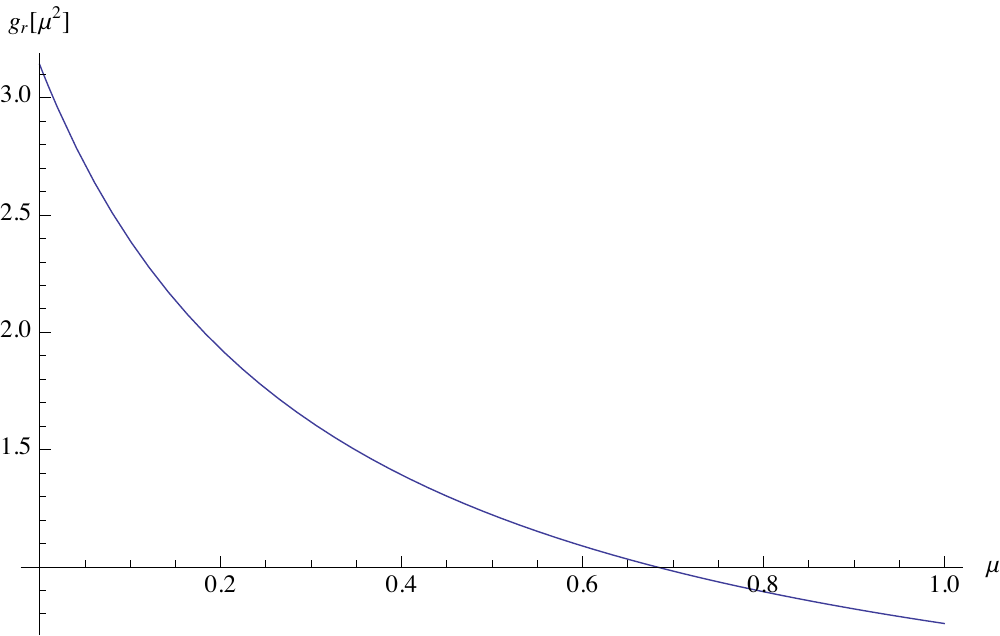}
\par\end{centering}
\caption{ Running of the dimensionless coupling constant $g_r$ for $d=1$.  We have chosen $\lambda^2 {\cal B} =1$. }
\label{grun1}
\end{figure}

This can also be written as
\bq
g_r[\mu^2] = \frac{g_r[\mu_0^2] } {1+ g_r[\mu_0^2] J[d] (\mu^{d-2} -  \mu_0^{d-2} )/(2-d) }.
\eq
In two dimensions one needs to  use Eq. \eqref{expand} to extract the finite logarithmic behavior of the running coupling constant.  For $d \neq 2$ in the dimensional regularized theory, both $\lambda$ and $\lambda_r$ are finite. 
In the LOAF approximation we find for $d<2$ an infrared stable fixed point at $g^\star$
\bq
g^\star = (2-d)/J[d].
\eq. 
We find in $d=1$, 
 $ g_r[\mu=0] = g^\star = \pi $.  Explicitly we have in one dimension 
 \bq
 \lambda_r^2[\mu^2] = \frac{\lambda^2 \mu \pi } {\mu \pi+ {\cal B} \lambda^2}.
 \eq
We see that
$\lambda^ 2= \lambda_r^2[\mu^2 \rightarrow \infty]$
For the dimensionless coupling constant we have 
 \bq
 g_r[\mu^2] =\frac{\lambda^2  \pi {\cal B}  } {\mu \pi+ {\cal B} \lambda^2}.
 \eq
  One get the behavior for $g_r[\mu^2] $ shown in Fig. \ref{grun1}.

\begin{figure}
\begin{centering}
\includegraphics[width=0.8\columnwidth]{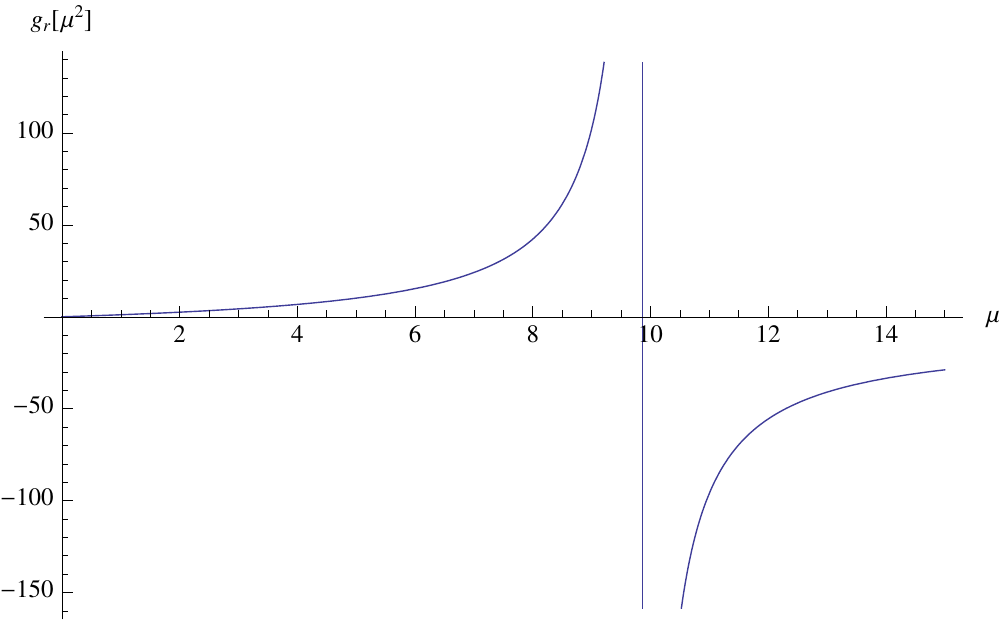}
\par\end{centering}
\caption{ Running of the dimensionless coupling constant $g_r$ for $d=3$.  We have taken   ${\cal B} \lambda^2  =1$}.
\label{grun3}
\end{figure}

 For $d>2 $,  $ g_r[\mu=0] =0$, and $I[d] < 0$.  Thus there is also a maximum value
of $\mu$ that one can reach before one hits the Landau pole at $1+ g_0 [\mu^\star] I[d]$.
 When d=3, we obtain
 \bq
\lambda_r^2 [\mu^2 ] =   \lambda^2   \left( 1-  \mu  \frac{\cal B }{ \pi ^2 }  \lambda^2   \right) ^{-1}.
\eq
We find now that  $\lambda_r^2 [\mu^2=0 ]= \lambda^2$ and 
The dimensionless renormalized coupling is given by 
\bq
g_r [\mu^2]=  {\cal B}  \lambda_r ^2[ \mu^2 ]  \mu=  \mu \lambda^2 {\cal B}   \left( 1-  \mu  \frac{\cal B }{ \pi ^2 }  \lambda^2   \right) ^{-1}.
\eq
and is clearly zero at $\mu =0$.  It  grows until it hits the Landau pole at  
\bq
1=  \frac{{\cal B}  \lambda^2   \mu}{\pi^2}.
\eq
This is shown in  fig. \ref{grun3}, where we have chosen ${\cal B} \lambda^2  =1$.

\begin{figure}
\begin{centering}
\includegraphics[width=0.8\columnwidth]{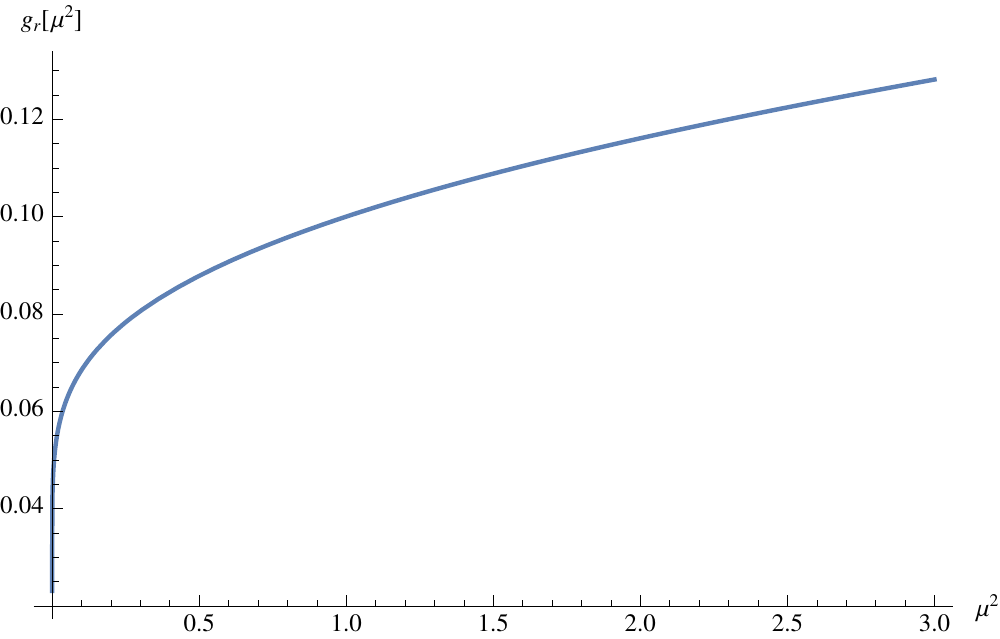}
\par\end{centering}
\caption{ Running of the dimensionless coupling constant $g_r$ for $d=2$, choosing $g_r[\mu_0^2]= 1/10, \mu_0 =1$ }
\label{run2}
\end{figure}

In two dimensions we have to take the limit of Eq. \eqref{vpp}  as $d \rightarrow 2$ to obtain 
\bq
g_r[\mu^2] = \frac{g_r[\mu_0^2] } {1-g_r[\mu_0^2] \ln [{\mu^2/\mu_0^2}]}
\eq
The behavior of $g_r[\mu^2]$ for small $\mu^2$ is displayed in Fig. \ref{run2}.

\section{Renormalized Effective Potential}
Once we have determined the renormalized running coupling constant we now use Eq. \eqref{vpp} for 
\bq
\frac{ \partial ^2  V_{eff}} {\partial \sigma ^2}= - \frac{1}{\lambda_r^2[\sigma/\nu^2] }  \label{m2} 
\eq   
to evaluate the renormalized effective potential by integrating back up. In two dimension, which is the critical dimension, we have 
integrating \eqref{m2}  once with respect to  $\sigma$  
\bq
\frac{\partial V}{\partial \sigma} =-\frac{\sigma}{\lambda_r^2[\mu^2]  }+\frac{v^2}{2}  + \frac{\calA}{32 \pi \nu^3}   \sigma  \left(\log \left(\frac{\sigma }{\nu^2  \mu^2}\right)-1\right),
\eq
which when set equal to zero yields what is known as the gap equation for $\sigma$.
Integrating once more with respect to $\sigma$  we obtain for the renormalized effective potential 
\ba
V_{eff}[v, \sigma]&&= -\frac{\sigma ^2}{2 \lambda_r^2[\mu^2] }+\frac{v^2}{2} \sigma \nonumber \\
&& +   \frac{\calA}{32  \pi  \nu^3} \sigma^2 \left[\frac{1}{2}  \log \left(\frac{\sigma }{\nu ^2 \mu^2}\right)-\frac{3 
   }{4} \right] \label{v2d}.
   \ea
We need to evaluate the effective potential at the point where 
$\sigma$ is the solution to the gap equation
\bq
\frac{\sigma}{\lambda_r^2[\mu^2]  }= \frac{v^2}{2}  + \frac{\calA}{32 \pi\nu^3}   \sigma  \left(\log \left(\frac{\sigma }{\nu^2  \mu^2}\right)-1\right).
\eq

In terms of $\sigma [v] $ evaluated using the gap equation we have the simple expression:
\bq
V_{eff}[\sigma, \mu ]= \frac{\sigma ^2}{2 \lambda^2 [\mu^2] }
 -\frac{\calA \sigma ^2}{32 \pi \nu ^3}  \left[ \log {\frac{\sigma }{\mu ^2 \nu ^2}}-\frac{1}{2} \right].
   \eq
   
\begin{figure}
\begin{centering}
\includegraphics[width=0.8\columnwidth]{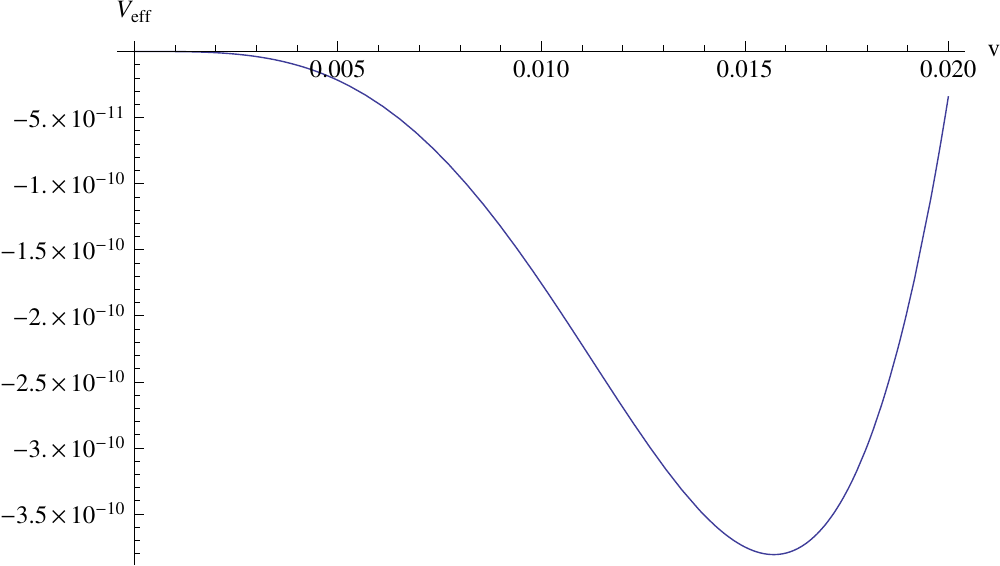}
\par\end{centering}
\caption{ Effective Potential in leading order in ${\cal A}$  approximation for $d=2$ using $\lambda = \mu = \nu =1$. and ${\cal B}=1/10$ }
\label{2dpot0}
\end{figure}

\begin{figure}
\begin{centering}
\includegraphics[width=0.8\columnwidth]{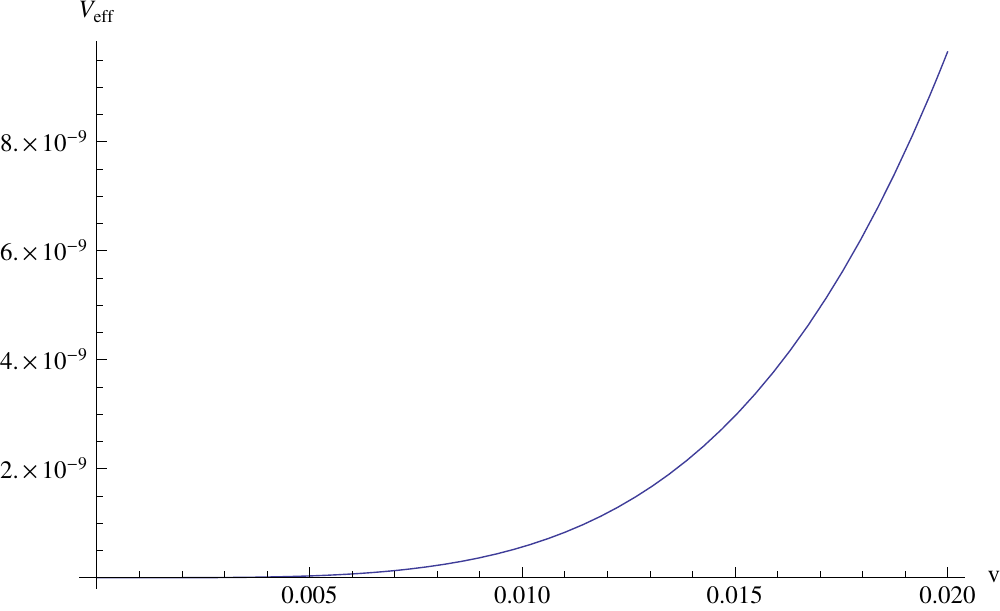}
\par\end{centering}
\caption{ Effective Potential in the LOAF   approximation for $d=2$ using $\lambda = \mu = \nu =1$. and ${\cal B}=1/10$ }\label{2dpot}
\end{figure}

 When we expand $V_{eff} $ in a power series in ${\cal B}= \calA/ (32 \pi \nu^3) $ we obtain to first order
   \bq
   V_{eff}[ v] =\frac{\lambda^2 v^4}{8} + \frac{\lambda^4 v^4 {\cal B} }{8}  \left( \log
   \left(\frac{ \lambda^2 v^2}{2 \mu^2 \nu^2}\right)-3/2 \right).
   \eq
   We notice that the leading term in the correction to the classical answer has opposite signs for the LOAF approximation and the first term in an expansion in $\calA$ (the loop expansion discussed by Hochberg et. al. \cite{physica}).  Thus even small correction of order $\calA^2$  as evidenced here can change the character of the answer.  The loop expansion for the effective potential as well as the Hartree approximation for the effective potential lead to a double well structure such as that seen in fig. \ref{2dpot0}.  On the other hand the LOAF approximation does not exhibit spontaneous symmetry breakdown as shown in fig \ref{2dpot}.  
   
To make contact with previous work of Hochberg et. al. \cite{physica}  we will re-obtain the RG equation for the renormalized coupling from the effective potential.
We will obtain the same RG equation for $\lambda_r[\mu^2]$ as that obtained in their paper.  
The fact that the potential cannot depend on the scale  where one defines the renormalized coupling constant leads to the renormalization group  equation:
\bq
\mu \frac {d V_{eff} }{d \mu} =0.
\eq
From Eq. \eqref{v2d} we then obtain
\bq
 \tilde \beta_{\lambda } = \mu  \frac{d \lambda_r[\mu^2] }{d \mu} = \frac{ \calA \lambda_r^3[\mu^2]} {16  \pi \nu^3} ,
 \eq
which agrees with our previous calculation Eq. \eqref{betalam} once  we compare it with  the slightly different definition of $\beta$ found in \cite{physica}.  However we must point out that our calculation, as distinct from the one in \cite{physica},  is non perturbative in $\calA$.  Also because of our use of dimensional regularization we are able to determine the $\beta$ function as a function of the dimension $d$ (see Eq. \eqref{betalam}). 

When we are not in the critical dimension  $d=2$, 
we can  integrate Eq. \eqref{vpp} with $\mu^2 =0$  once  with respect to $\sigma$.
 \bq
\frac {\partial V_{eff}}{\partial \sigma} = - \frac{\sigma}{\lambda^2} + \frac{1}{\lambda} ( \frac{\lambda v^2}{2} + f_0) + c + \frac{ \calA  } { \nu (2-d) 4 d } J[d] [ \frac{\sigma}{\nu^2}] ^{d/2}.
  \eq
  Here $\lambda = \lambda_r[\mu=0]$. 
One can chooses the constant of integration $c$ so that  
\bq
c+ f_0/\lambda = f_r/\lambda.
\eq
We then have
\bq
\frac {\partial V_{eff}}{\partial \sigma} =  - \frac{\sigma}{\lambda^2} +  \frac{ v^2}{2}+\frac{f_r}{\lambda } - \frac{\calA  } {(2-d) 4 d   \nu } J[d] [ \frac{\sigma}{\nu^2}] ^{d/2}   \label{vp}.
  \eq
  The gap equation is
 \bq
  \frac{\sigma}{\lambda^2} =  \frac{ v^2}{2}+ \frac{f_r}  {\lambda}  - \frac{\calA  } {(2-d) 4 d   \nu } J[d] [ \frac{\sigma}{\nu^2}] ^{d/2}.   \label{gapr}
 \eq 
 If we want to choose the renormalized theory to correspond to the massless KPZ equation one then chooses $f_r=0.$
 Note that Eq. \eqref{gapr}  is similar to the gap equation in the large-$N$ expansion for a relativistic  $\phi^4$ \cite{CJP}  theory.  Integrating \eqref{vp} once with respect to $\sigma$  we obtain
 \bq
V_{eff} [\sigma, v] =  - \frac{\sigma^2}{2 \lambda^2} +  \sigma \frac { v^2}{2}  - \frac{\calA } {(2-d)  \nu }\frac { \sigma }{ 2 d (2+d) } J[d] [ \frac{\sigma}{\nu^2}] ^{d/2}   \label{veff2}.
\eq

To obtain $ V_{eff} [v] \equiv V_{eff} [\sigma[v], v]$ one needs to solve the gap equation \eqref{gapr}  for $\sigma[v]$.
Note that the condition for being at a minimum in $V_{eff} [v]$ , is just
\bq
\frac{dV_{eff}[v]}{dv} = \frac {\partial  V_{eff} }{\partial v}|_\sigma +  \frac {\partial  V_{eff} }{\partial \sigma}|_v \frac{ \partial \sigma[v]}{\partial v} = \frac {\partial  V_{eff} }{\partial v}|_\sigma =0,
\eq
since we are evaluating the potential at the solution of the gap equation.  Therefore in the LOAF approximation 
the condition that we are at a minimum is 
\bq
\frac{dV_{eff} [v]}{dv} = \sigma v =0.
\eq
If there is broken symmetry at this order ($ v \neq 0$), this can only happen when $\sigma =0$.  From the gap equation Eq. \eqref{gapr}, we see if $f_r=0$ then of necessity $v=0$ and there is no broken symmetry.  This condition will be loosened if we calculate to next order.  This is exactly what happens in the $1/N$ calculation of the BEC phase transition \cite{Baym}. 

\subsection{1d}
In 1d  the potential is given by 
\bq
V_{eff} [\sigma, v] =  - \frac{\sigma^2}{2 \lambda^2} +  \sigma \frac { v^2}{2}  - \frac{\calA } {6 \nu^2 \pi }\sigma^{3/2}  \label{veff1d} ,
\eq
where $\sigma[v]$ is the solution of the gap equation
 \bq
\sigma= \lambda^2 \left[   \frac{ v^2}{2}  -  \frac{\calA  } {4 \nu^2  \pi}  \sigma^{1/2}  \right] \label{gap1d}.
 \eq
Note that to leading  order in  ${\cal A}$ we obtain
 \bq
V_{eff} [ v] =   \lambda^2 \frac { v^4}{8}  - \frac{\calA } { 6 \nu^2 \pi }( \frac{\lambda^2 v^2}{2})^{3/2}  \label{veff1dold}.
\eq
This is exactly the same answer as Eq. 42 in \cite{physica}. To leading order in $\calA$  in the fluctuations there is double well behavior.
However once we add a little bit of second order effects, this seems to vanish, at least in the LOAF approximation. 

\begin{figure}
\begin{centering}
\includegraphics[width=0.8\columnwidth]{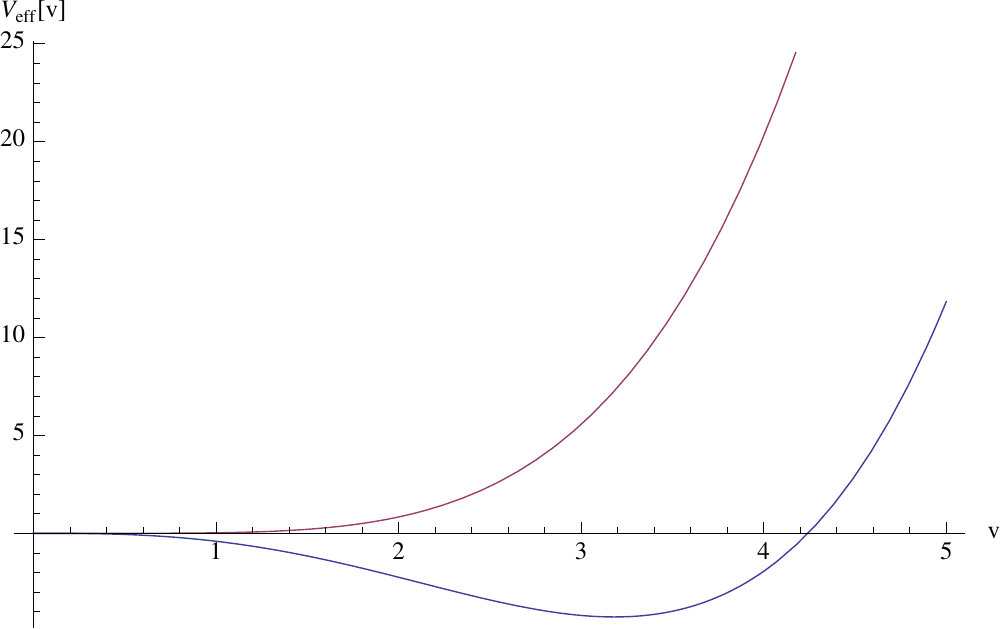}
\par\end{centering}
\caption{ Effective Potential for $d=1$ choosing values $\lambda = b=1$. The upper curve is the LOAF result Eq. \eqref{vloaf1}, the lower curve is the double well solution found
at one loop (Eq. \eqref{veff1dold})}
\label{fig:1d}
\end{figure}

In one dimension we can explicitly solve the gap equation for $\sigma[v]$, to obtain
\bq
\sigma^{1/2} = \sqrt{ \lambda^2 \frac { v^2}{2}  +b^2/4} - b/2,
\eq
where 
\bq 
b= \frac{ \lambda^2 \calA}{ 4 \nu^2 \pi}.
\eq
Thus when we keep the self consistent corrections we get:
\begin{widetext}
\bq
V_{eff} [v] = -\frac{\left(\sqrt{\frac{b^2}{4}+\frac{\lambda ^2 v^2}{2}}-\frac{b}{2}\right)^4}{2
   \lambda ^2}-\frac{2 b \left(\sqrt{\frac{b^2}{4}+\frac{\lambda ^2
   v^2}{2}}-\frac{b}{2}\right)^3}{3 \lambda ^2}+\frac{1}{2} v^2
   \left(\sqrt{\frac{b^2}{4}+\frac{\lambda ^2 v^2}{2}}-\frac{b}{2}\right)^2 \label{vloaf1}.
\eq
\end{widetext}
 This is strictly monotonically increasing in $v$, and contains terms to all orders in $\calA$.  Of course the  leading order in $\epsilon$ solution cannot have a broken symmetry solution since
 one cannot satisfy the broken symmetry condition $\sigma=0$ except at $v=0$.  The potential is quite flat in $v$ as seen in the upper curve in Fig. \ref{fig:1d}. The lower
 curve is the one loop result which suggests a broken symmetry solution. 

 \subsection{ three spatial dimension ($d=3$)}

From Eq. \eqref{veff2}  setting $d=3$ we get 

\begin{figure}
\begin{centering}
\includegraphics[width=0.8\columnwidth]{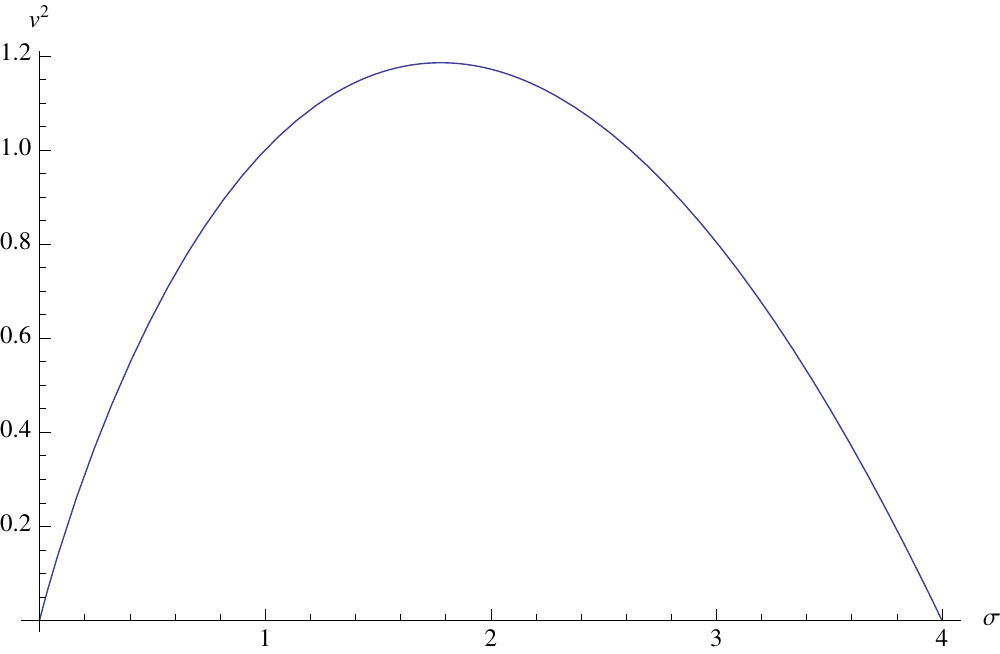}
\par\end{centering}
\caption{ $v^2$ as a function of $\sigma$ using $\lambda = \nu= \calA=1 $.}
\label{vsq3}
\end{figure}

\bq
V_{eff} [\sigma, v] = 
 - \frac{\sigma^2}{2 \lambda^2} +  \sigma \frac { v^2}{2}  +  \frac{\calA } {30  \nu^4 \pi^2 } \sigma^{5/2}  \label{veff3d}.
\eq
The gap equation is now
\bq
\sigma = \lambda^2 \left[  \frac{v^2}{2}  +  \frac{\calA } {12  \nu^4 \pi^2 }\sigma^{3/2}   \right] ,
\eq
or equivalently
\bq
v^2 =\frac{2 \sigma }{\lambda ^2}-\frac{\calA \sigma ^{3/2}}{6 \pi ^2 \nu ^4} \label{gap3}.
\eq
We notice that now $v^2$ is a parabolic like function of $\sigma$ which rise to a maximum and then falls to zero.
This is shown in fig. \ref{vsq3}.  Eq. \eqref {gap3} can be explicitly solved for $\sigma$ as a function of $v^2$, to reproduce the first half of the parabola. However it is sufficient for our purpose to just rewrite $V_{eff}$ in terms of $\sigma$ since at small  $\calA, \lambda$  the relationship between $\sigma$ and $v^2$ is  linear over a wide range of $v^2$. 

\begin{figure}
\begin{centering}
\includegraphics[width=0.8\columnwidth]{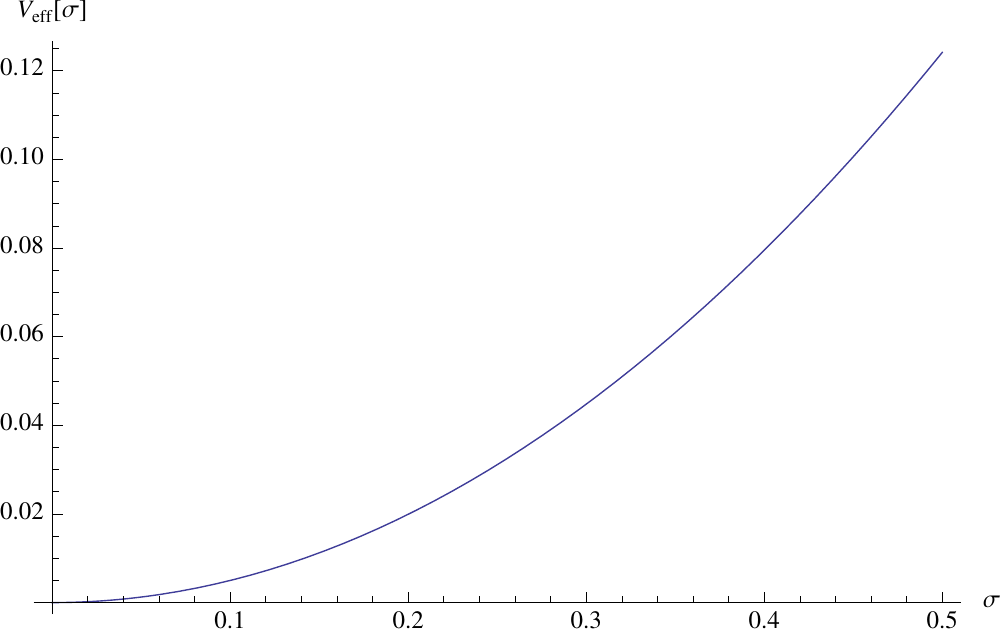}
\par\end{centering}
\caption{ Effective Potential $ V[\sigma, v^2[\sigma] ] $with $v^2$  evaluated using the gap equation for  $d=3$. Here  $\lambda = \nu =\calA=1$.}
\label{v3d1}
\end{figure}

The value of $\sigma$ at the maximum $\sigma =\frac{64 \pi ^4 \nu ^8}{\calA^2 \lambda ^4}$, which leads to maximum value of $v^2 = \frac{64 \pi ^4 \nu ^8}{\calA^2 \lambda ^4}$.  For values of $v^2$ below the maximum  $v^2$ is a monotonically increasing function of $\sigma$ and we can rewrite
$V$  simply as a function of  $\sigma$, using the gap equation \eqref{gap3}: 
\bq
V[\sigma] = \frac{\sigma ^2}{2 \lambda ^2}-\frac{\calA \sigma ^{5/2}}{20 \pi ^2 \nu ^4} \label{v3d}.
\eq
This is plotted in fig. \ref{v3d1}.

If we reexpand Eq. \eqref{v3d} in terms of $v^2$ up to order $\calA$ we obtain the one loop result:
\bq
V_{1-loop} = \frac{\lambda ^2 v^4}{8}+ \frac{\calA \lambda ^5 v^5}{120 \sqrt{2} \pi ^2 \nu ^4},
\eq
which is Eq. (105) in \cite{Hochberg} and also does not display symmetry breaking.  In the self consistent Gaussian approximation discussed in \cite{Gaussian}, the authors found a double well structure in three spatial dimensions which differs from the one loop result and our result (which preserves Goldstone's theorem). 

\section{Comparison with Exact Results}
For the LOAF approximation
\bq
\beta(g_r) = \mu \frac{\partial}{\partial \mu} g_r =  (d-2)g_r + J(d) g_r^2  \label{beta}.
\eq
In particular
\begin{equation}\label{KPZ.e:RGF-d13-beta-II}
   \beta[ \, g_r \,]
   =
   \begin{cases}
      - g_r \, ( \, 1 - g_r / \pi \, ) \>,
      \qquad
      &
      \text{for $d=1$, and} 
      \\
       g_r \, ( \, 1 + g_r / \pi^2 \, ) \>,
      \qquad
      &
      \text{for $d=3$.} 
   \end{cases}
\end{equation}
Thus we see that if $d<2$ there is a fixed point at 
\bq
g^\star = \frac{(2-d) } {J(d) }.
\eq
For $d>2$ LOAF predicts that there is only the fixed point at $g^\star=0$.  This is in contradistinction to the
 exact result for the $\beta$ function for $d>2 $, which  was obtained by  Lassig \cite{Lassig}.  The exact result was found  by mapping the KPZ problem into a directed polymer problem. That is for $d>2$ one can make the Cole-Hopf transformation
\bq
\phi = \frac{2 \nu} {\lambda}  \ln w,
\eq
and obtain the differential equation for $w$
\bq
{\dot w} = \nu \nabla^2 w + \frac{\lambda}{2 \nu} w \eta. 
\eq
This equation can be derived from the Martin Siggia Rose Lagrangian  \cite{ref:GN}
\bq
L =  \int dx \left[ {\tilde w} \left( {\dot w} - \nu \nabla^2 w -\frac{\lambda}{2 \nu} w \eta \right) \right].
\eq
One can now perform the integration over $\eta$ for white noise to obtain the effective action
\bq
\int d^d x dt [ \left[ {\tilde w} \left( {\dot w} - \nu \nabla^2 w \right)  -\frac{\lambda^2 \calA }{2 \nu^2 } {\tilde w}{\tilde w} w w \right]. \label{polymer}
\eq
This is quite similar to the action for the annihilation problem $A+A \rightarrow 0$ except that the force in Eq. \eqref{polymer} is attractive.  For that problem the only renormalization is the due to summing the  loop corrections to the scattering  which renormalizes the effective bare coupling
$
 {\tilde \lambda }  = \frac{\lambda^2 \calA }{2 \nu^2 }.
$
 Note now  the effective  bare coupling parameter depends also on $\calA$ and $\nu$, the quantities that are necessary to renormalize in the original formulation of the
 KPZ equation to obtain the correct renormalized result.
Summing the chain of bubble diagrams one finds
 \bq
 \tilde \lambda_r [\mu^2] = Z_g[\mu^2] \tilde \lambda;, ~~ Z_g^{-1} = 1-\frac{\mu^{d-2} \tilde \lambda  B[d] }{\nu (2-d)}.
 \eq
 where 
\bq
 B[d] = 2! 2^{-d/2} (4 \pi) ^{-d/2} \Gamma(2-d/2).
 \eq

 One then obtains the  exact RG equation for the dimensionless coupling constant   
 \bq g_r = \frac{ \tilde \lambda_r   }{\nu} \mu^{d-2}.  \eq
 
 \bq
 \beta (g_r) = \mu {\partial g_r}{\partial \mu} = g_r (d-2) - B[d] g_r^2  \label{exactch}.
 \eq
 Here $B[d] = 2 2^{-d/2} (4\pi) ^{-d/2}  \Gamma[ 2- d/2] $. 
 This leads to the fact that there is an unstable  fixed point point for $d>2$ with 
 \bq
 g_r^\star = (d-2)/B[d].
 \eq

The fact that we do not find this unstable fixed point for $d>2$  in leading order LOAF using the KPZ action is due to the fact that we have not included the $\mu$ dependence of  the variables
$\calA$ and  $\nu$.  These effects come from vacuum diagrams that are of  order $\calA^2$ in the loop expansion, and of order $\epsilon$ in the auxiliary field expansion. 
Standard calculations in perturbation theory show that the renormalization of $\nu$  changes the sign of the last term  in  Eq. \eqref{beta}.  In a subsequent paper \cite{us}  we will show that one can do a LOAF expansion for the Cole-Hopf transformed KPZ  action given by  Eq. \eqref{polymer}, and get a $\beta$ function which qualitatively agrees with the exact result Eq. \eqref{exactch}.

\begin{figure}
\begin{centering}
\includegraphics[width=0.8\columnwidth]{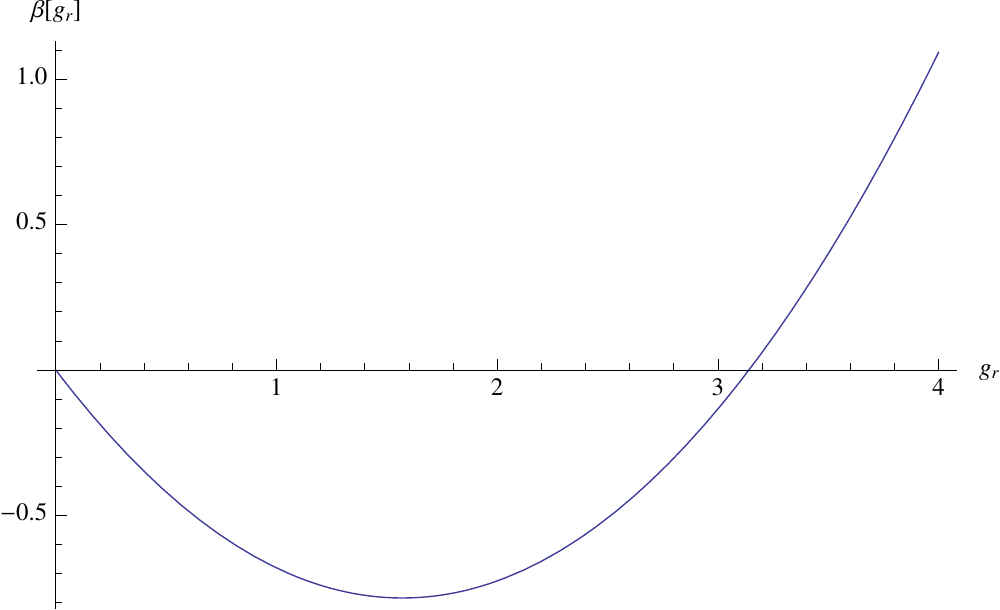}
\par\end{centering}
\caption{$\beta$ as a function of $g$ for $d=1$ in LOAF.}
\label{flow}
\end{figure}

Summarizing our results pictorially, using  the original form of the KPZ action in the Onsager Machlup form we find that in 
 LOAF the flow  $\beta (g,d=1)$ in leading order  is shown in fig.  \ref{flow}. For $d>2 ,  \beta$ is a monotonically increasing function of $g_r$ and the only fixed point is at the origin
$g_r=0$. 

\begin{figure}   
\begin{centering}
\includegraphics[width=0.8\columnwidth]{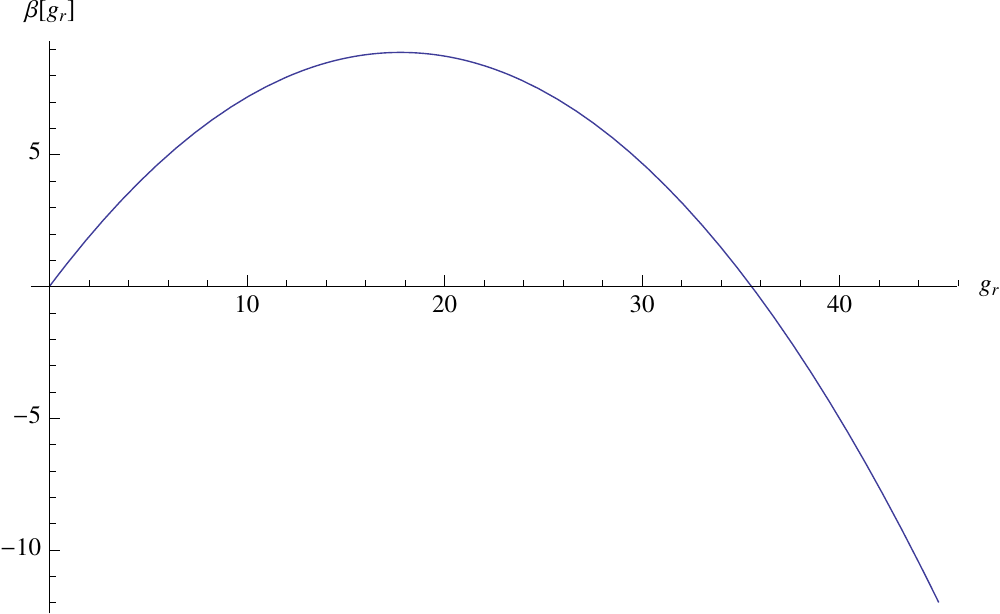}
\par\end{centering}
\caption{Exact  $\beta$ function as a function of $g_r$ for $d=3$.}
\label{exact}
\end{figure}

This is to be contrasted with the the results of Lassig \cite{Lassig} (and also the results found by us using the Cole-Hopf transformed action and the LOAF approximation \cite{us} )  which predicts an unstable fixed point and a roughening transition for $d>2$ as shown in fig. \ref{exact}.

\section{Discussion and Conclusions}
In this paper we have given the general approach for obtaining the effective action for noisy reaction diffusion equations. We have presented a strategy for performing 
a non perturbative (in noise strength and coupling constant)  re-summation of the theory. In leading order the approximation is similar to a Hartree approximation which assumes only Gaussian fluctuations for the field $\phi$. However our approximation preserves Ward identities order by order in the expansion parameter $\epsilon$ so that it is more trustworthy when discussing symmetry breakdown.   In our approach we are able to obtain not only an expression for the effective potential, but also  the effective action to any order in the  auxiliary field loop expansion parameter $\epsilon$.  Starting from our expression for the effective action one could then derive the time evolution equation for the average value of the field $\phi$ where the average is over various runs with Gaussian noise, as well as equations for the noise generated correlation functions. 
In obtaining the effective potential we did not need to resort to introducing a background field such as that used in \cite{physica} or  \cite{Gaussian}. 
Our  lowest order approximation (LOAF), when reexpanded as a series in $\calA$  reduced  to the results  of the loop expansion in the noise strength $\cal A$  \cite{physica}. Although that is formally true,  at small $\calA$ our results differed qualitatively  from the result found in \cite{physica}  in that there was no symmetry breaking found in any dimension.  The one loop approximation suggested that there is dynamical symmetry breaking in both one and two dimensions. 
 The Gaussian effective potential discussed in \cite{Gaussian} , which is related to the Hartree approximation leads to an analytic expression for the effective potential that is similar in structure to that found here.  In that approximation there is also an auxiliary field $\Omega$ which is similar to $\sigma$ which is self-consistently determined.  However the Hartree approximation does not preserve Goldstone's theorem in general \cite{Andersen} and thus the result that it predicts symmetry breaking for the massless KPZ equation is suspect.  In the LOAF approximation, the inverse of the auxiliary field propagator which is related to the second derivative of the effective potential is a renormalization group invariant.  From that we derived the running of the coupling constant $\lambda$.  In the LOAF approximation for the effective potential we were unable to determine the renormalization of  $\nu$ and $\calA$.  To obtain the full RG equations using our approach one must go beyond the LOAF approximation. One must either seperately evaluate the momentum dependent Green's functions and vertices   or do a gradient expansion of the effective action and write  the leading order terms in the form  discussed in Kardar and Zee \cite{Kardar-Zee}. 
Namely to obtain the renormalization of $\nu$ and $\calA$, they show it is sufficient to only look at terms in the gradient expansion of the  action which are quadratic in $\phi$.  The general form of these terms
are
\bq
S_{eff} =  \int dx dt \left( \alpha \frac{\partial \phi } {\partial t} - \beta \nabla^2  \phi) \right)^2.
\eq
These correction terms get generated by momentum dependent corrections to the $\phi \phi$ correlation function that occur in next order in the LOAF expansion. 
  We will bypass this complication in a subsequent paper by
using the LOAF approximation for the Cole-Hopf transformed KPZ action  in the MSR formulation of the action \cite{us}.  There by introducing additional composite fields we will be able to obtain an RG flow in LOAF that is qualitatively correct.

\appendix
 \section{Cutoff Approach to Renormalization}
 
In this appendix we will introduce an explicit momentum dependence so we can see the relationship between the bare and renormalized parameters explicitly. 
In general we have: 
\bq
V_{eff} [\sigma, v] =  - \frac{\sigma^2}{2 \lambda^2} + \frac{\sigma}{\lambda} ( \frac{\lambda v^2}{2} + f_0) +  \frac{\calA \nu }{2} K[\frac{\sigma}{\nu^2}] ,
\eq
where
\bq
K[ m^2 ] =  \int_0^\Lambda   \frac{d^d k}{(2 \pi)^d} |k | (\sqrt{k^2 +  m^2}.
\eq
The ultraviolet divergent terms are isolated (except in 2d where one has to be careful about infrared issues) by recognizing that the integrand has the following large $k$ behavior. 
\bq
|k|   \sqrt{k^2 +  m^2} = k^2 + \frac{m^2}{2} - \frac{m^4}{8k^2} + \ldots  \label{expansion}
\eq

\subsection{one dimension}
In one dimension one needs to add and subtract the first two terms of Eq. \eqref{expansion} to obtain :
\ba
V_{eff} [\sigma, v] &&=  - \frac{\sigma^2}{2 \lambda^2} + \frac{\sigma}{\lambda} ( \frac{\lambda v^2}{2} + f_0)  \nonumber \\ 
&&+  \frac{\calA \nu }{2} \left[\Lambda^3/3 + \frac{\sigma}{\nu^2}\Lambda  - \frac{1}{3 \pi} (\frac{\sigma}{\nu^2})^{3/2}  \right]. 
\ea
Thus we identify the first divergent term in the brackets as an infinite constant shift in the vacuum (or cosmological term in field theory jargon).  The second term is the renormalization of $f$.  That is we define
\bq
\frac{f_r}{\lambda} = \frac{f_0}{\lambda} + \frac{\calA \Lambda}{2 \nu} ,
\eq
and then set $f_r = 0$ so since we are considering the "massless" KPZ equation. 
This leaves 
\bq
V_{eff} [\sigma, v] =  - \frac{\sigma^2}{2 \lambda^2} + \frac{\sigma}{\lambda} ( \frac{\lambda v^2}{2} ) - \frac{\calA \nu }{6 \pi} (\frac{\sigma}{\nu^2})^{3/2}  
\eq

\subsection{2 dimensions} 
In two dimensions we have  
\bq
K[ m^2 ] =  \frac{1}{2 \pi}   \int_0^\Lambda  k^2 \sqrt{k^2 +  m^2}.
\eq
A direct expansion for large $k$ has the problem of infrared divergences, which we will regulate by introducing an infrared mass $\mu$.  We will handle the ultraviolet  divergences  then by adding and subtracting the following terms:
\ba
K[ m^2 ] &&= K_{reg} [m^2, \mu^2]  \nonumber \\
&&+    \frac{1}{2 \pi}  \int_0^\Lambda   ( k^3 + \frac{ k m^2} {2} -  \frac{ m^4 k^2 }{8 (k^2 + \mu^2)^{3/2}  }) \label{kmsq}, \nonumber \\
\ea
where the regulated  finite integral is 
\ba
&&K_{reg} [ m^2 , \mu^2] =    \nonumber \\
&&\frac{1}{2 \pi} \int_0^\Lambda  \left ( k^2 \sqrt{k^2 +  m^2}   - k^3 - \frac{ k m^2} {2} + \frac{ m^4 k^2 }{8 (k^2 + \mu^2)^{3/2}  } \right) \label{K2d}. \nonumber \\
\ea
Explicitly  
\bq
K_{reg} [ m^2, \mu^2] =   \frac{m^4 }{16  \pi} \left(  \frac{1}{2}  \ln \frac{m^2}{\mu^2} - \frac{3}{4} \right).
\eq
The divergent part of $K$ is 
\ba
&&K_{div}  [ m^2, \mu^2] =   \frac{1}{2 \pi}\left[ \frac{\Lambda^4}{4}-\frac{1}{8} m^4  \log
   \left(\sqrt{\frac{\Lambda^2+\mu^2}{\mu^2}}+\frac{\Lambda}{\mu}\right) \right. \nonumber \\
   && \left. +\frac{\Lambda^2 m^2 }{4} \right].
\ea
The first term is related to vacuum renormalization, the second to a renormalization of the coupling constant and the third renormalizes  $f_0$.
The first term is an irrelevant constant, the second we will discuss below.  The renormalization of $f$ is the equation for the coefficient of the term linear in $\sigma$.
We have
\bq
f_r/\lambda_r = f_0/\lambda + \frac{\calA \Lambda^2}{16 \pi \nu}.
\eq
Since we are interested in the massless KPZ equation, after renormalization we set $f_r=0$.  To  see that the second term is related to coupling constant renormalization
we define the renormalized coupling constant from the second derivative of the effective potential with respect to $\sigma$. Namely
 \ba
\frac{1}{\lambda_r^2[\sigma/\nu^2] } && = -  \frac{ \partial ^2  V } {\partial \sigma ^2} \nonumber \\
&& = \frac{1}{\lambda^2}+  \frac{\calA  } {8 \nu^3 } \int_0^\Lambda 2 \pi k^2   \frac{d k}{(2 \pi)^2} \frac{1}{ ( k^2 + \frac{\sigma}{\nu^2})^{3/2}}   \label{l2d}. \nonumber \\
\ea
Defining a coupling constant at the arbitrary  mass scale $\mu^2$, we have 
 \bq
\frac{1}{\lambda_r^2[\mu^2] }  = \frac{1}{\lambda^2}+  \frac{\calA  } {8 \nu^3 } \int_0^\Lambda   k^2   \frac{d k}{(2 \pi) }\frac{1}{ ( k^2 + \mu^2)^{3/2}}. \label{lmu2d}
\eq
Note that the last term in Eq. \eqref{lmu2d}   is exactly the last term one has in the divergent part of $K[\mu^2]$, namely the last term in Eq. \eqref{kmsq}.  
In terms of this coupling constant, the coupling constant at $\sigma/\nu^2$ is now finite. 
\bq
\frac{1}{\lambda_r^2[\sigma/\nu^2] } = \frac{1}{\lambda_r^2[\mu^2] }   + \frac{\calA}{32 \pi \nu^3}  \ln{ \frac{\mu^2 \nu^2}{\sigma}}. \label{ug}
\eq
This coupling constant grows logarithmically until it hits the Landau pole which can be demonstrated by rewriting Eq. \eqref{ug}
in the form
\bq 
\lambda^2[\sigma/\nu^2]  = \frac{\lambda^2 [\mu^2]}  { 1- \frac{\calA}{32 \pi \nu^3}   \lambda^2[\mu^2]  \ln{ \frac {\sigma} {\mu^2 \nu^2}}}.
\eq
Integrating Eq. \eqref{l2d}   once with respect to $\sigma$ we obtain  the gap equation:
\bq
\frac{\partial V}{\partial \sigma} =-\frac{\sigma}{\lambda_r^2[\mu^2]  }+\frac{v^2}{2}  + \frac{\calA}{32 \pi \nu^3}   \sigma  \left(\log \left(\frac{\sigma }{\nu^2  \mu^2}\right)-1\right).
\eq
Integrating once more we obtain for the potential 
\ba
V_{eff}[v, \sigma]&&= -\frac{\sigma ^2}{2 \lambda_r^2[\mu^2] }+\frac{v^2}{2} \sigma \nonumber \\
&& +   \frac{\calA}{32 \pi \nu^3} \sigma^2 \left[\frac{1}{2}  \log \left(\frac{\sigma }{\nu ^2 \mu^2}\right)-\frac{3 
   }{4} \right]. \nonumber \\
   \ea
   \subsection {three dimensions} 
\bq
V_{eff} [\sigma, v] =  - \frac{\sigma^2}{2 \lambda^2} + \frac{\sigma}{\lambda} ( \frac{\lambda v^2}{2} + f_0) +  \frac{\calA \nu }{2} K[\frac{\sigma}{\nu^2}],
\eq
where in three dimensions
\bq
K[ \mu^2 ] =  \int  \frac{d^3 k}{(2 \pi)^3} |k | (\sqrt{k^2 +  \mu^2} - \sqrt{k^2}].
\eq
and we have subtracted a vacuum energy term.  Introducing a cutoff $\Lambda$  we have that 
\bq
K[\mu^2] = \frac{1}{4 \pi ^2} \left[ \mu^2 \frac{\Lambda^3}{3} - \mu^4 \frac{\Lambda}{4} + \frac{4}{15} \mu^5 \right].
\eq
Thus we see we get an infinite contribution to the mass as well as to the inverse $\sigma$ propagator which we need to interpret properly.
We have
\ba
V_{eff}[v, \sigma]&&=\frac{f \sigma }{\lambda }-\frac{\sigma ^2}{2 \lambda ^2}+\frac{\sigma 
   v^2}{2} \nonumber \\
&&+ \frac{\calA \nu} {8 \pi
   ^2} \left(\frac{\Lambda^3 \sigma }{3 \nu ^2}-\frac{\Lambda \sigma ^2}{4 \nu
   ^4}+\frac{4}{15} \left(\frac{\sigma }{\nu ^2}\right)^{5/2}\right) \nonumber \\
      \ea
   Taking one derivative we get
   \bq
  \frac{\calA \Lambda^3}{24 \pi ^2 \nu }-\frac{\calA \Lambda \sigma }{16 \pi ^2 \nu ^3}+\frac{\calA \sigma
    \sqrt{\frac{\sigma }{\nu ^2}}}{12 \pi ^2 \nu ^3}+\frac{f}{\lambda
   }-\frac{\sigma }{\lambda ^2}+\frac{v^2}{2}.
   \eq
   We see that the tadpole term  $ f $ gets renormalized
 \bq
   \frac{f_r}{\lambda} = \frac{f}{\lambda}  +  \frac{\calA \Lambda^3}{24 \pi ^2 \nu }.
\eq
To study the massless KPZ equation one needs to set $f_r =0.$.  Taking a second derivative of the partially
renormalized potential one gets:
\bq
 \frac{\partial^2 V}{\partial \sigma^2} \equiv - \frac {1} {\hat{\lambda}^2[\sigma]} =-\frac{\calA L}{16 \pi ^2 \nu ^3}+\frac{\calA \sqrt{\sigma }}{8 \pi ^2 \nu
   ^4}-\frac{1}{\lambda ^2}.
   \eq
We next define a  finite renormalized coupling constant at the scale  $\sigma/\nu^2 = \mu^2 $
 \bq
  \lambda_r^2 [\mu^2]  = {\lambda}_r ^2 [\sigma/\nu^2 = \mu^2 ] ,
  \eq    
  so we can write
   \bq
 \frac{\partial^2 V}{\partial \sigma^2} \equiv - \frac {1} {\lambda_r^2[\sigma/\nu^2 ]} =-  \frac {1} {\lambda_r^2 [\mu^2 ] } +\frac{\calA }{8 \pi ^2 \nu^3}
 \left( \sqrt{ \sigma/\nu^2 } - \sqrt{\mu^2} \right)  \label{run3d}.
   \eq
Having performed these renormalizations we can integrate back up to get the renormalized gap equation and renormalized effective potential.

\begin{acknowledgments}

I would like to thank Juan Perez-Mercader for thoughtful discussions and suggesting this research. I  would also like to thank Gourab Ghoshal and John Dawson for  valuable discussions. 

\end{acknowledgments}


\begin{thebibliography}{9}

\bibitem{Itzykson} C. Itzykson and J.-B. Zuber, "Quantum Field Theory"  McGraw Hill, New York (1980). 
 \bibitem{Hochberg} D. Hochberg, C. Molina-Paris, J. Perez-Mercader, M. Visser,  Phys.Rev.  {\bf E60},  6343 (1999).
 \bibitem{Dodd} P. J. Dodd and N. M. Ferguson PLoS One {\bf 4} (9); e6855. doi:10.1371/journal.pone.0006855
\bibitem{Zorzano} D. Hochberg and M.P. Zorzano, Physica A {378}, 238 (2007). 

 \bibitem{Andersen}  J. Andersen, Rev. Mod. Phys. {\bf 76}, 599 (2004). 
\bibitem{us} F. Cooper  "Auxiliary Field Loop expansion for the Effective Action for Stochastic Partial Differential equations II", Harvard Preprint.

 \bibitem{BCG} C. Bender, F. Cooper, and G. Guralnik Annal Phys {\bf 109}, 165 (1977).

 \bibitem{BEC1} F.  Cooper,Chih-Chun Chien,B. Mihaila, J. F. Dawson, and E. Timmermans, Phys. Rev. Lett. {\bf 105} 240402 (2010). 
 \bibitem{BEC2} F.  Cooper, B. Mihaila, J. F. Dawson, Chih-Chun Chien and E. Timmermans, Phys. Rev. A  {\bf 83} 053622 (2011). 
\bibitem{BCS} C.A. R. Sa de Melo, M. Randeria, and J. R. Engelbrecht, Phys. Rev. Let. {\bf  71}, 3202 (1993). 
\bibitem{Hubbard} J. Hubbard, Phys. Rev. Lett. {\bf 3} , 77 (1959), R. L. Stratonovich, Doklady  {\bf} 2, 416 (1958).

\bibitem{CJP} S. Coleman, R. R. Jackiw, and H. D. Politzer, Phys. Rev. D {\bf 10} 2491 (1974). 




\bibitem{SD1} F. Cooper, B. Mihaila, and J. F. Dawson Phys. Rev. D. {\bf 70}, 105008 (2004). 
\bibitem{Goldstone} J. Goldstone, Nuovo Cimento, {\bf 19}, 154 (1961). 
\bibitem{Onsager}L. Onsager, S. Machlup  Phys. Rev. {\bf}  91, 1505 (1953).
\bibitem{Graham} R. Graham, "Springer Tracts on Modern Physics {\bf 66}", Springer, Berlin (1973). 
\bibitem{Zinn-Justin} J. Zinn-Justin, Nucl. Phys. B {\bf 275} (FS17) ,135 (1986)
\bibitem{msr} P.C. Martin, E. Siggia, and H. Rose, Phys. Rev. {\bf A8}, 423, (1973).
\bibitem{Peliti} L. Peliti, J. Physique {\bf 46} 1469 (1985).
\bibitem{janssen} H. P. Janssen , Z. Physik {\bf B 23} , 377 (1976). 
\bibitem{Jouvet} B. Jouvet, R. Phythian, Phys Rev. {\bf A19} 1350 (1979).


\bibitem{chiral}  A. Chodos, F. Cooper, W. Mao and A. Singh, Phys.Rev.D {\bf 63} 096010 (2001). 
\bibitem{Chien} Chih-Chun Chien and F.  Cooper,  Phys.  Rev.  {\bf A 87} , 045602 (2013).

\bibitem{Nreview} Moshe Moshe and  Jean Zinn-Justin,  Phys. Rept. { \bf 385} , 69 (2003).
\bibitem{bcf} C. Bender, F. Cooper and B. Freedman, Nucl. Phys. B {\bf 219}, 61 (1983).

\bibitem{ref:GN}J. Cardy "Field Theory and Non-Equilibrium Statistical Mechanics" Lectures presented as part of the Troisieme Cycle de la Suisse Romande, Spring 1999.
\bibitem{Kamenev} A. Kamenev "Field Theory of Non-Equlibrium Systems" Cambridge Press (2011).

\bibitem{kpz} M. Kardar, G. Parisi, and Y-C Zhang, Phys. Rev. Lett. {\bf 56},889 (1986).

\bibitem{physica} D. Hochberg, C. Molina-Paris, J. Perez-Mercader, and M. Visser , Physica A {\bf 280} 437 (2000).
\bibitem{Gaussian}    F.S. Amaral, I. Roditi, Physica A {\bf 385} 137 (2007).

\bibitem{SK} J. Schwinger, J. Math. Phys. 2, 407 (1961),~~
L.V. Keldysh, Zh. Eksp. Teor. Fiz. {\bf 47} (1964). 

\bibitem{ref:Hochberg1} D. Hochberg, F. Lesmes, F. Moran, and J. Perez-Mercader, Phys. Rev. E. {\bf 68}, 0066114 (2003).




\bibitem{Zanella} J. Zanella and E. Calzetta, Phys. Rev. E {\bf 66}, 036134 (2002) 

\bibitem{Baym} G. Baym, J.P. Blaizot, M. Holzmann, F. Laloe, and D. Vautherin, Phys. Rev. Lett. {\bf 83}, 1703 (1999). 




\bibitem{Lassig} Michael Lassig, Nucl. Phys. B {\bf 448} [FS] 559, (1995). 
\bibitem{review} Uwe Tauber, Martin Howard and Benjamin P Vollmayr-Lee, J. Phys. A:Math.Gen. {\bf 38} R79 (2005).
\bibitem{Kardar-Zee} M. Kardar and A.  Zee , Nuclear Physics B  {\bf 464}  [FS]  449 (1996).
\bibitem{Lee}
B.~P. Vollmayr-Lee, J. Phys. A: Math. Gen. {\bf 27},  2633  (1994).


\end{thebibliography}
\end{document}